\documentclass[a4paper,fleqn,usenatbib,useAMS]{mnras}

\pdfoutput=1
\ifpdf
\usepackage{graphicx}
\usepackage[latin1]{inputenc}
\usepackage{color}
\usepackage{times}
\usepackage{natbib}
\usepackage{setspace}
\usepackage{amsmath}
\usepackage{amsfonts}
\usepackage{amssymb}
\usepackage{multirow}
\usepackage{multicol}
\usepackage{aas}
\usepackage{booktabs}
\usepackage{adjustbox}
\usepackage{bm}		
\usepackage{pdflscape}	
\usepackage[T1]{fontenc}
\usepackage{ae,aecompl}

\usepackage{newtxtext,newtxmath}

\usepackage{epstopdf}
\usepackage{subfigure,lscape,color}
\newif\ifAMStwofonts
\AMStwofontstrue
\definecolor{red}{rgb}{1,0.,0.}

\title[The SFR-$M_{\star}$ and SSFR-$M_{\star}$ relations in local star-forming galaxies.]{Morphology rather than environment drives the SFR-Mass relation in the local universe}

\author[R. Calvi]{R. Calvi$^{1,2}$\thanks{Contact e-mail: \href{mailto:rcalvi@iac.es}{rcalvi@iac.es}}, B. Vulcani$^{3,4}$, B. M. Poggianti$^{4}$, 
A. Moretti$^{4}$, J. Fritz$^{5}$, G. Fasano$^{4}$
  \\
  $^{1}$Instituto de Astrof{\'{\i}}sica de Canarias, E-38200 La Laguna, Spain \\
  $^{2}$ Departamento de Astrof{\'{\i}}sica, Universidad de La Laguna, E-38205 La Laguna, Spain\\
  $^{3}$ School of Physics, University of Melbourne, VIC 3010, Australia \\
  $^{4}$INAF - Astronomical Observatory of Padova, 35122 Padova, Italy\\
  $^{5}$Instituto de Radioastronom\'\i a y Astrof\'\i sica, IRyA, UNAM, Campus Morelia, A.P. 3-72, C.P. 58089, Mexico
}

\date{}

\pubyear{2018}

\begin{document}
\label{firstpage}
\pagerange{\pageref{firstpage}--\pageref{lastpage}}
\maketitle

\begin{abstract}

Exploiting a sample of 680 star-forming galaxies from the Padova-Millennium Galaxy
Group Catalog (PM2GC) \citep{Calvi11} in the range $0.038\leq z \leq 0.104$, we present a detailed
analysis of the Star Formation Rate (SFR)- stellar mass ($M_{\ast}$) and specific SFR(SSFR)-$M_{\ast}$
relations as a function of environment. We adopt three different parameterizations of environment,
 to probe different scales. We consider separately 1) galaxies in groups, binary
and single systems, defined in terms of a Friends-of-Friends algorithm, 2) galaxies located
at different projected local densities, 3) galaxies in haloes of different mass. Overall, above
log$M_{\ast}/M$>10.25 and SSFR>$10^{-12}yr^{-1}$, the SFR-$M_{\ast}$ and SSFR-$M_{\ast}$ relations do not depend
on environment, when the global environment is used, while when  the halo mass is considered, high mass haloes might have a systematically lower (S)SFR-$M_{\star}$ relation. Finally, when local densities are exploited, at any given mass galaxies in less dense environments have systematically  higher values of SFR. All the relations are characterized by a large scatter ($\sigma \sim 0.6$), which is due
to the coexistence of galaxies of different morphological types. Indeed, at any given mass,
late-types are characterized by systematically higher values of SFR and SSFR than S0s and
ellipticals. Galaxies of the same morphology show similar trends in all the environments, but
their incidence strongly depends on environment and on the environmental parametrization
adopted, with late-types generally becoming less common in denser environments, contrasted
by the increase of ellipticals and/or S0s. Our results suggest that in the local universe morphology and local interactions, probed by the local density parameterization,
 have dominant roles in driving the characteristics of the SFR-$M_\ast$
relation.
\end{abstract}


\section{Introduction}
Galaxy populations have been shown to evolve differently over the last 8 Gyr:  more massive galaxies formed most of their stars rapidly at $z>1$. Observationally, it has been found that the bulk of the star formation has occurred between redshifts $\rm z\sim 1.5-2$. At lower redshifts,  the cosmic Star Formation Rate (SFR) density declines by a factor of 10  \citep{Madau96,Cowie96,Lilly96,Madau98,Hogg98,Flores99,Haarsma00,Hopkins04} and the most vigorous sites of star formation activity are lower mass systems (the {\it downsizing} effect, \citealt{Gavazzi02,Boselli06}). 

 The histories of star formation are strongly reflected in galaxy colors. Although some peculiar galaxies show a large scatter in color, the great majority of normal galaxies form two populations closely related with the Hubble types: the ``red  sequence'' galaxies are mostly passive and massive with, in prevalence, an early-type morphology, while the ``blue cloud'' galaxies  are typically less massive, star-forming and with a late-type morphology \citep{Strateva01,Kauffmann03,Baldry04,Balogh04}.

These two populations give a different contribution to the decline of the SFR. Indeed, the number and the stellar mass density of  red galaxies have grown by a factor of 2 from $\rm z\sim 1$ to 0, while those of  blue galaxies have remained nearly constant \citep{Vandenbosch02}. This implies that the decline of global Star Formation History (SFH) is due both to the fact that  most of the star formation in the local Universe occurs at a more moderate rate in star-forming galaxies than in the past, and mainly to the fact that currently there are more red galaxies than in the past, suggesting that a fraction of blue galaxies have stopped forming stars and contributed to the growth of the red sequence \citep[e.g.][]{Guglielmo15}.
\\

The suppression of the star formation in blue galaxies appears to be accelerated in denser environments, where the fraction of star forming galaxies at fixed stellar mass and redshift is lower compared to the field \citep{Kauffmann04,McGee11, Guglielmo15}. In the most massive structures the fraction of early type galaxies increases, due to an increase of lenticular galaxies to the detriment of blue star-forming spiral ones \citep{Butcher84,Aragon93,Dressler97,Fasano00,Lewis02,Gomez03,Treu03,Balogh04,Postman05,Smith05,Desai07}, suggesting that some cluster-specific physical processes  drive the evolution of the galaxy properties. Mechanisms such as ram-pressure and/or tidal stripping \citep{Gunn72}, galaxy-galaxy interactions \citep{Kewley06,Sobral15,Stroe15} and ``harassment'' \citep{Moore96} are the most accredited to act on star-forming disk galaxies that fall down into the clusters, leading both a decline in the star formation activity and a morphological transformation. 

A way to understand how galaxies evolve and the physical processes at the origin of the quenching is to investigate how the SFR and the SSFR (the star formation rate per unit stellar mass) of star-forming galaxies of a given mass vary in different environments.   A tight relation between SFR and stellar mass for star-forming galaxies (the so-called star formation ``main sequence'', MS) has been observed in the field both in local samples \citep{Noeske07a,Zamojski07,Zheng07,Karim11} and at higher redshifts \citep{Daddi07,Elbaz07,Pannella09,Wuyts11} where the relation shifts to higher SFR at given mass. As predicted by semi-analytic models \citep{Finlator06,Finlator07}, the steady growth of SFR with mass is a consequence of the smooth growth of SFHs with time, caused by both the gas infalling from the intergalactic medium (IGM) and feedback effects. 

To date, very few works focused on the variation of the SFR-mass relation with the environment. \citet{Peng10}, analysing samples of star-forming  galaxies drawn from the SDSS and  zCOSMOS \citep{York00,Lilly07}, showed that both the SFR-$M_{\star}$ and SSFR-$M_{\star}$ relations do not depend on the local density. In contrast, several works  showed that, at a given galaxy stellar mass, the star formation activity of galaxies in groups and clusters is on average reduced with respect to the field,  both above and below $z\sim 1$ \citep{Patel09,VonderLinden10,Haines13,Vulcani10,Lin14,Allen16,Paccagnella16}.

The SFR has also a strong dependence on galaxy morphology \citep{Kennicutt92}, thus the environmental dependence of the SFR-$M_{\star}$ relation could be a dependence on galaxy morphology too. The  environment is usually more relevant for lower mass galaxies, than for more massive ones \citep[e.g.][]{Vulcani15,Kravtsov04,Zolotov12, Kenney14}. Evidence for the suppression of SFR in cluster galaxies compared to the field galaxies with the same morphological type has been found by many authors \citep{Gisler78,Kennicutt83,Dressler85,Balogh98,Hashimoto98,Yuan05,Bamford09}, though some works also showed that the interaction of late-type cluster galaxies with their environment results in enhanced star formation \citep{Kennicutt84,Gavazzi85,Moss93}. \citet{Deng10} found that galaxies of the same type have  higher SFRs and SSFRs at lower local densities and that the environmental dependence of the SFR-$M_{\star}$ and SSFR-$M_{\star}$ relations of late-type galaxies is stronger than that of early-type galaxies. More recently \cite{Abramson14}, re-normalizing the SFR by disk stellar mass, argued that the SF efficiency does not depend on galaxy mass for star-forming disks, while \citet{Willett15} found no statistically significant difference in the SFR-$M_{\star}$ relations in a wide range of nearby morphological sub-types of star-forming disk galaxies, suggesting that these systems are strongly self-regulated. 

In this work we present an homogeneous and complete study of  the dependence of the SFR-$M_{\star}$ and SSFR-$M_{\star}$ on the environment and morphology in the local universe. 
Our goal is to shed light on the mechanisms that drive the suppression of SFRs. We  will study these relations using different parametrizations of environment. We will adopt a ``global environment'' parametrization, i.e. considering  the  large scale and  subdividing galaxies in single, binary systems and groups, a ``local environment'' parametrization, i.e. taking into account the number of neighbors around each galaxy, and consider the ``mass of the hosting halo'', obtained from simulations. These three different definitions are most likely related to different physical processes.   

The paper is structured as follows. In \S2 we describe the data samples and the galaxy properties. \S3 includes a brief explanation of the different methods used to parametrize the environment.  \S4 presents the results of our analysis, performed considering the global environment, local densities and halo masses and the galaxy morphologies. In \S5 we summarize our findings and conclude. Throughout this paper we adopt the $\Lambda$CDM cosmology with $H_{0}=70\, \rm km \, s^{-1} \, Mpc^{-1}$, $\Omega_{m}=0.3$ and $\Omega_{\Lambda}=0.7$.

\section{The DATA-SET }

We base our analysis on the Padova Millennium Galaxy and Group Catalogue  \citep[PM2GC,][]{Calvi11}. The catalog was extracted from the Millennium Galaxy Catalogue (MGC), a B-band imaging and deep survey which consists of 144 overlapping fields covering a total area of $\rm \sim$37.5 deg$^{2}$ \citep{Liske03,Driver04} fully contained within the regions of both the Two Degree Field Galaxy Redshift Survey (2dFGRS) and the Sloan Digital Sky Survey (SDSS). Details on the catalog can be found in \cite{Calvi11}. Briefly,   galaxies in the range 0.03$\leq z\leq$0.11 were selected from the MGCz catalogue, the spectroscopic extension of MGC-BRIGHT catalogue, which contains  objects 
with B$_{MGC}$$<$20mag with a $\rm \sim$96\% spectroscopic completeness. 
The PM2GC  contains 3210  galaxies with an absolute magnitude brighter than M$_{B}$=-18.7, corresponding to the K-corrected B$_{MGC}$=20 magnitude at our redshift upper limit. 
Applying a Friends of Friends (FoF) algorithm, 
catalogues of groups, binary systems and isolated galaxies were identified (see Sec.3.1 for details about the FoF).

Galaxy stellar masses were computed following the \cite{BelldeJong01} relation, which relates the Mass-to-light ratio and galaxy color:
$\log(M_{\star}/L_B)=-0.51+1.45\times(B-V)$.
Rest frame (B-V) color were computed from the SDSS Galactic extinction-corrected model magnitudes in $g$ and $r$. All masses were then scaled from a \cite{Salpeter55} to a \citet{Kroupa01} Initial Mass Function (IMF), adding -0.19 dex to the logarithmic value.
 These mass estimates 
 are in  good agreement  with those obtained with SDSS-DR7 for PM2GC \citep{Calvi11}. The dispersion is similar to the typical mass error of $\rm \sim$0.2-0.3dex   \citep[see, e.g.,][]{Kannappan07}.

\begin{figure}
\centering
 \includegraphics[scale=0.36]{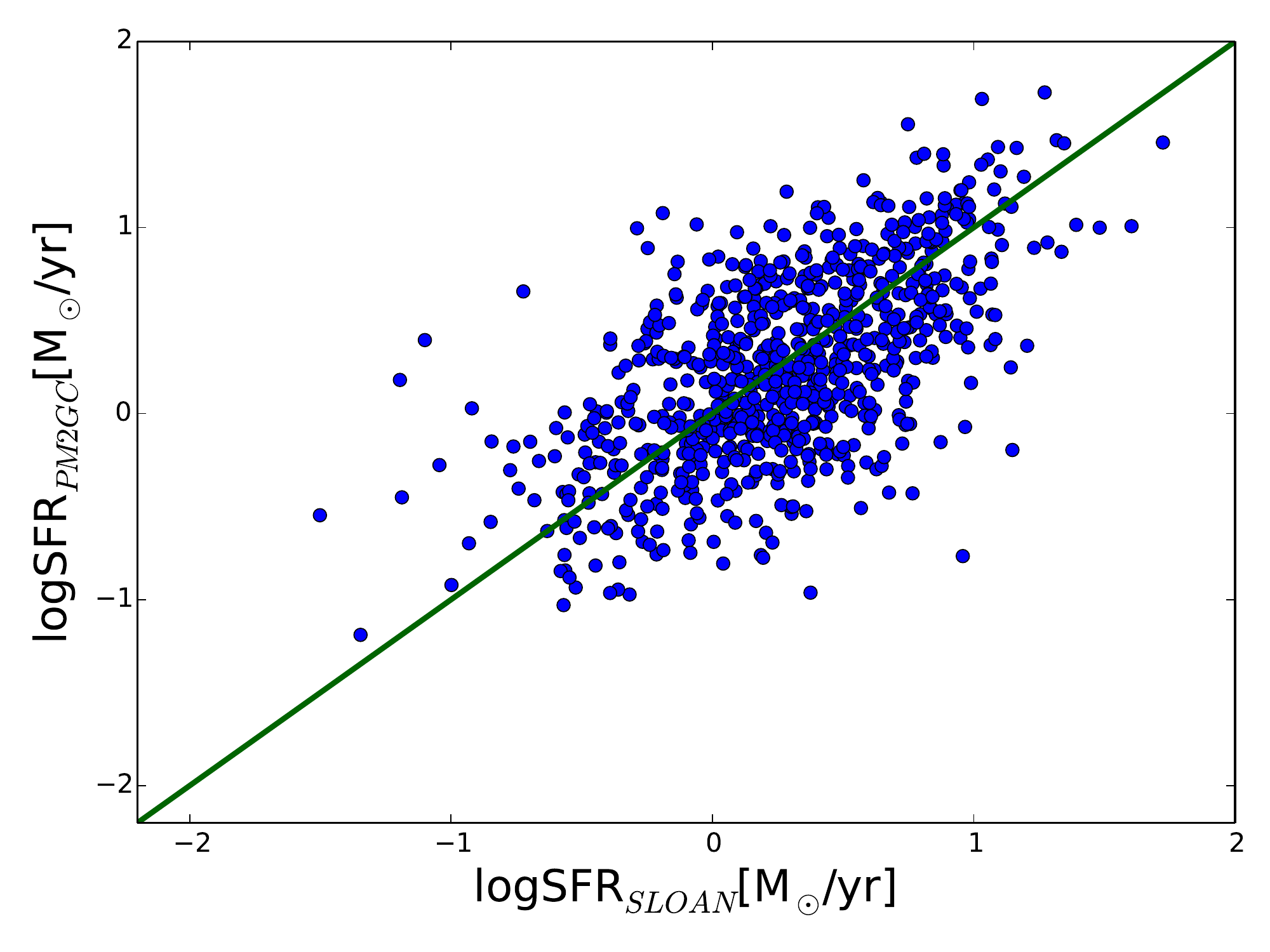}
\caption{Comparison between the SFRs computed by SINOPSIS and those released by SDSS-DR4  for the common sample of 778 galaxies. The green line is the 1:1 relation.}\label{SFconfr}
\end{figure}

SFR estimates have been presented in \citet{Poggianti13}. They were  computed using  SINOPSIS, an automatic tool  based on a stellar population synthesis technique that reproduces the observed optical spectra of galaxies through the fitting of the stellar content, extinction and, when possible, metallicity. The model is fully described in \citet{Fritz07,Fritz11,Fritz14}. A set of template spectra with the main features of an observed spectrum, i.e equivalent widths of emission and absorption lines and fluxes emitted in the continuum, are fitted summing the synthetic spectra of Simple Stellar Population (SSP) of different ages. The SSPs used are taken from a combination of two different sets that both use Padova evolutionary tracks \citep{Bertelli94}. The first set uses the \citet{Jacoby84} library of observed stellar spectra in the optical domain. The second set of SSPs uses the MILES library \citep{Sanchez04,Sanchez06}. 

PM2GC galaxies have SDSS, 2dFGRS and MGCz spectra \citep{Driver05}, the latter taken with the same instrument and setup of the 2dFGRS. We always chose the highest quality spectrum available to feed SINOPSIS. SDSS spectra were used in 86\% of the cases,  2dFGRS spectra in 12\% and MGCz spectra in the remaining cases. SDSS spectra are flux-calibrated. For 2dF spectra a relative flux calibration was performed using the response curve provided by the 2dF team and a refined calibration, using a curve derived by the ratio between the 2dF spectra and the corresponding SDSS spectra for objects with both spectra available, following the procedure described in \citet{Cava09}. A similar approach has been followed also for the MGCz spectra.

SFRs were computed using the sets of Simple Stellar Populations (SSPs) built with a standard \cite{Salpeter55} Initial Mass Function (IMF) in the range 0.15-120 $M_\odot$. They have then been  converted to a \cite{Kroupa01} IMF. 
As most of the PM2GC spectra are provided by SDSS, we tested the reliability of the spectrophotometric technique by comparing the estimates of SFR obtained by SINOPSIS to those provided by \cite{Brinchmann04} for a subsample of 778 objects, matched in coordinates
within 2 arcsec.
We considered only SDSS galaxies classified as star forming or low S/N star forming. Figure \ref{SFconfr} shows a good correlation for the majority of galaxies in the two samples, with a dispersion of $\sim$0.4 dex.

Morphologies were determined using MORPHOT \citep{Fasano12}, an automatic tool designed to reproduce as closely as possible the visual classifications. MORPHOT adds to the classical CAS (concentration/asymmetry/clumpiness) parameters  \citep{Conselice03} a set of additional indicators derived from digital imaging of galaxies and has been proved to give an uncertainty only slightly larger than the eyeball estimates. It was applied to the B-band MGC images to identify ellipticals, lenticulars (S0s), and later-type galaxies \citep{Calvi12}.

\subsection{The galaxy sample}

The global PM2GC sample consists of 3210 galaxies with M$_{B}<-18.7$, but the number of galaxies with an output from SINOPSIS is 3160. We excluded from this sample active galactic nuclei (AGN) as in principle their SFR might be overestimated because the emission line fluxes are considered as produced only by star formation processes, and not by other mechanisms. To exclude  AGNs, we rely on the AGNs classification of the SDSS\footnote{http://www.sdss3.org/dr10/spectro/spectro\_ access.php}. We matched the SDSS and PM2GC catalogues  within 5 arcsec and excluded galaxies identified as ``broad-line''. 57 galaxies in the general field are AGN.

We limit our analysis to the sub-sample complete  both in stellar mass and SFR in the redshift range 0.038$\leq z\leq$0.104 (see also Sec. 3.1), which is the same redshift range of group galaxies.

The stellar mass completeness limit was computed as the mass of the reddest galaxy at the upper redshift limit. The mass of the reddest $\rm M_{B} =-18.7$ galaxy (B-V =0.9) at z=0.11 is equal to $\rm M_{\star} = 10^{10.25}M_{\odot}$ \citep{Calvi11}.  

The limit in SFR is dictated by the fact that below a threshold of 2\AA{} the emission measurements for any line (including $\rm H\beta$) in our spectra are considered unreliable 
\citep{Fritz14}. This sets a lower detection limit that in terms of specific star formation rate is SSFR=SFR/M$_\ast$=$10^{-12.5} yr^{-1}$. To be conservative, we  consider for the analysis only galaxies with current SSFR$>10^{-12} yr^{-1}$.

The total number of galaxies used in the analysis is 680.

\section{Measurements of environment}

As discussed in detail in \citet{Muldrew12} and \citet{Fossati15} , there is no universal environmental measure and the most suitable method generally depends on the spatial scale(s) being probed. 
For example, some definitions are more sensitive 
to the richness of the systems which they belong to,  some others to galaxy interactions. Here we  present the three different parametrizations we adopt to describe the environment in our analysis.

\subsection{Global environments}

According to the definition of global environment, galaxies are commonly subdivided into, for instance, superclusters, clusters, groups, the field and voids, which can be expected to roughly correspond to the galaxies' host halo mass.
For massive structures, memberships are based on redshifts and velocity dispersions of the galaxies. 

This definition is purely based on observational evidence, therefore it might be biased due to projection effects. 

At the group level, assigning membership is a  delicate task. In the literature, many algorithms exist to identify groups, but the most frequently applied is the Friends-of-Friends one \citep[see, e.g.,][for some published group catalogues]{Eke04,Berlind06}. This approach is based on a physical choice to link galaxies in groups. The FoF algorithm requires two different linking lenghts, one in radial velocity and one in projected distance on the plane of the sky.

For the PM2GC,  \cite{Calvi11} present the characterization of the group sample. They initially defined some trial groups linking galaxies within a line-of-sight depth equal to three times the velocity dispersion, fixed at 500 km/s rest frame, and a projected mutual separation of 0.5 h$^{-1}$Mpc \footnote{$h^{-1}=\frac{100}{H_{0}}$}.
Then, from the second iteration, they computed the velocity dispersions of these trial groups and restricted the membership within $\rm \pm 3\sigma$ from the median group redshift and $1.5 R_{200}$, with $R_{200}$ an approximation of the virial radius -
the radius which delimits a sphere with mean interior density 200
times the critical density - computed as in \citet{Finn05}
\begin{equation}
R_{200}=\frac{1.73\sigma}{1000kms^{-1} \sqrt{\Omega_{\Lambda}+\Omega_{0}(1+z)^{3}}}h^{-1}Mpc
\end{equation}
  176 groups with at least three members were identified, comprising in total 1057 galaxies in the range $0.04\lesssim z\lesssim 0.1$. The median velocity dispersion of groups is 192 km/s (the range is between $\sim$40 km/s and $\sim$1200 km/s). Non-group galaxies were subdivided into  ``field-single'' and ``field-binary''. 
The first sample  contains 1141 isolated galaxies with no companions within 0.5h$^{-1}$Mpc and 1500 km/s which from now on will be referred to as pure field; the second one is composed of 490 binary systems of galaxies, i.e. those pairs of bright galaxies that have a projected mutual separation less than 0.5h$^{-1}$Mpc and a redshift within 1500 km/s. 

Galaxies in groups, binary systems and singles, together with those galaxies that although located in a trial group, did not make it into the final group sample, are collected into the general sample.
In our analysis, we will consider only  groups with a velocity dispersion $\rm \sigma< 500km/s$, to eliminate a possible contamination from the clusters that might be possibly included in the PM2GC sample.  We have also evaluated how many binary system galaxies have a velocity dispersion $\rm \sigma > 1000 km/s$, to quantify the number of pairs which might be not properly bounded. This fraction is just the 5$\%$ of the entire binary sample, thus we decided to include also these galaxies to improve the statistic. From now on, we will indicate  as GS, GR, FB and FS the samples of all galaxies, groups, binary systems and single galaxies respectively.

\subsection{Local densities}
The local density is a proxy for the local environment, which is sensitive to the processes taking place on small scales (e.g. galaxy galaxy interactions, tidal interactions).

The local density can be parametrized in several ways, following different techniques. For example, it is possible to fix the metric aperture in which the number of neighbours of a galaxy are counted or to measure the distance to the $n_{th}$ nearest neighbour ($n$ in the range of 5-10).
Typically, inside a fixed aperture, the third- or fifth- nearest neighbor distance is used for small scales, while a larger number of neighbors provides better measurements of larger scale/denser environments.

The projected local galaxy density for PM2GC is derived calculating in a circular area A, projected on the sky, the N nearest galaxies brighter than an absolute V magnitude M$_{lim}^{V}$. The projected density is $\rm \Sigma = N/A$ in number of galaxies per $\rm Mpc^{2}$. For details see \citet{Vulcani12}. In brief, for each galaxy in the PM2GC, we have considered an area containing only the five nearest projected neighbours ($A_5$), to avoid the problem of too large a volume which could include galaxies belonging to other haloes. The absolute V magnitude has been computed from the absolute B magnitude provided by the MGC survey and the rest-frame index color $B-V$. Neighbours were counted within a fixed velocity difference $\rm \pm 1000$ km/s and only if brighter than M$_{lim}^{V}\leq$-19.85. 

Note that, due to the peculiar geometry of the area covered by the PM2GC survey (a stripe of 0.6 $\times$ 73 deg across the sky), the circular area $A_5$  used to compute local densities tends to extend off the survey area, thus producing increasingly unreliable estimates of the local density. To overcome this issue, \citet{Vulcani12} used the photometric and spectroscopic information for all galaxies in the regions of the sky around the MGC ($\rm \pm 1.5^{\circ}$) from the SDSS and 2dFGRS. These, together, yielded a highly complete sample in the regions of interest \citep{Vulcani12}.

\begin{figure}
\centering
\includegraphics[scale=0.4]{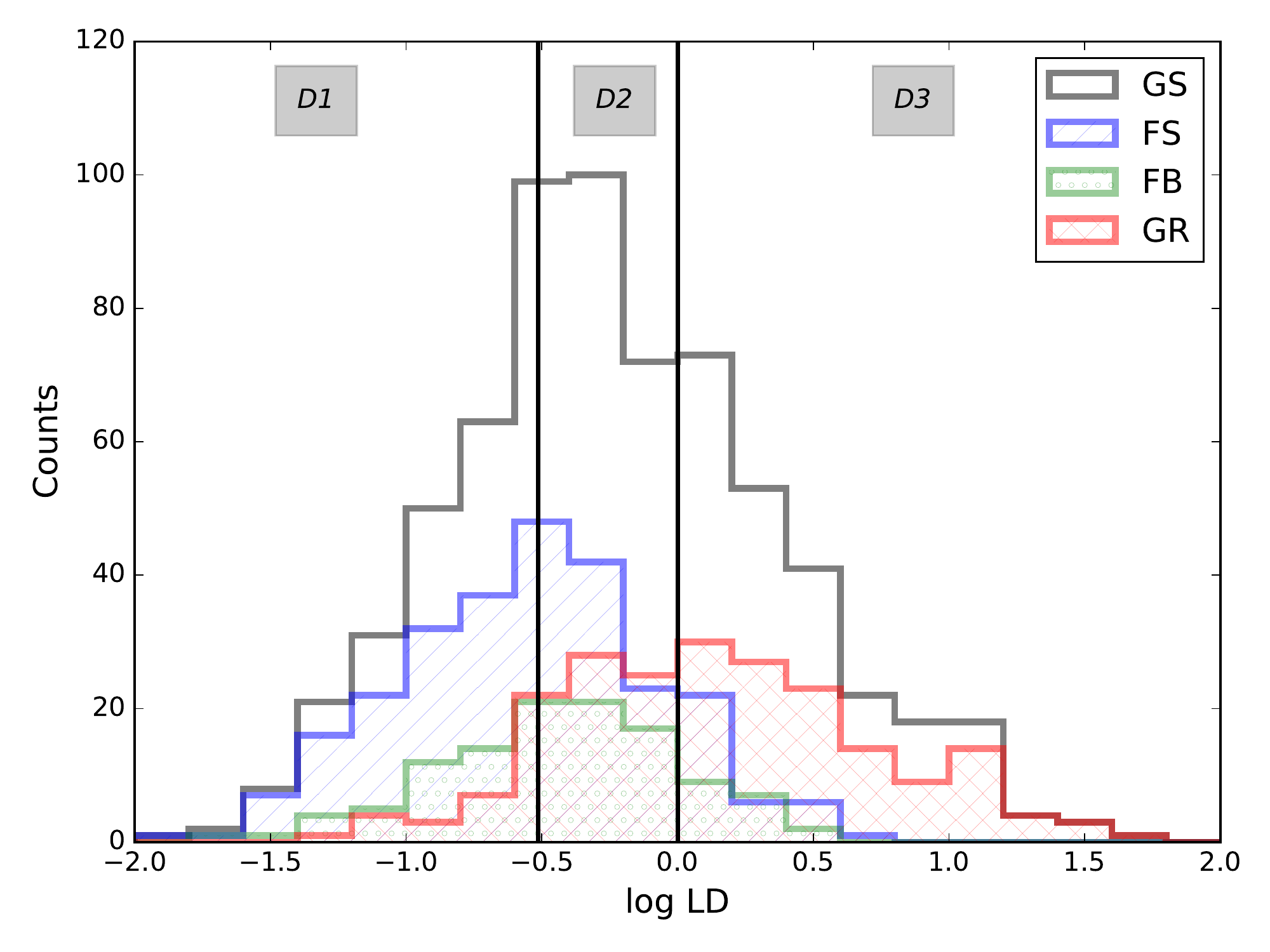}
\caption{Local density distributions of galaxies above the mass and SSFR limits in the different environments of PM2GC. The vertical lines represent the limits of the three density bins considered in Sec. 4. \label{hist_ld_1}}
\end{figure}

We stress that even if there is a sort of general correlation between global and local environments, the two definitions of environments are not at all equivalent \citep{Muldrew12}.
This is well visible in  Figure~\ref{hist_ld_1},  where  the distributions of local densities for the general sample, groups, binary, and single galaxies  are shown.  As also discussed by \citet{Vulcani12}, single galaxies are mostly located at lower densities, binaries at intermediate densities, while groups  at higher  densities, nonetheless there is a large overlap between the populations.  Very low density regions can be found even in groups. 

In what follows, we will consider galaxies belonging to three different bin of local density: 
D1: -2$\leq$LD$\leq$-0.5, D2: -0.5$<$LD$<$0, D3: 0$\leq$LD$\leq$1.8.

\subsection{Halo masses}
The mass of the dark matter halo hosting a galaxy is important for the evolution of that galaxy, so halo mass is an important environmental parameter.

Halo masses are generally computed exploiting simulations. 
\cite{Paccagnella2018}  developed  a functional formulation to compute halo mass estimates of the PM2GC structures, exploiting mock catalogs drawn from the Millennium Simulation \citep{Springel05}.
The method  is  based on the evidence that the mass of a dark matter halo associated with a group is tightly correlated with the total stellar mass of all member galaxies and to the mass of the most massive galaxy (MMG) \citep{Yang07,Yang08}.

Briefly, the starting point was to assign halo masses to galaxies in GR, FB and FS identified by the same FoF algorithm used on the observed catalogue \citep{Calvi11} in  10 simulated fields using only galaxies brighter than $\rm M_{B} < -18.7$. 
Halo masses were assigned to the sim-projected groups with at least three members. For each sim-projected group,  the simulated halo that includes the majority of its members and has the lowest level of contamination from interlopers was identified. The mass of such simulated halo was then associated to the sim-projected group.

A total stellar mass to halo mass relation was then built from simulations and applied to the PM2GC data to assign halo masses to the observed structures.

\cite{Paccagnella2018}  determined that our groups/binaries/singles span a halo mass range of $10^{11.6}-10^{14.8} M_\odot$ / $10^{11.2}-10^{12.9} M_\odot$ / $10^{10.9}-10^{13.9} M_\odot$, with a median value of $10^{12.68\pm 0.05} M_\odot$ / $10^{12.03\pm 0.03} M_\odot$ / $10^{11.58\pm 0.02} M_\odot$, respectively.

As for local density, we will analyze the trends dividing the samples in three bin of halo mass: MH1: 11.6$\leq M_{h}\leq 12.5$, MH2: $12.5< M_{h}<13.5$, MH3: 13.5$\leq M_{h}\leq 14.8$.

\section{Results}
We are now in the position of analyzing the SFR- and SSFR- $M_\ast$ relations both in the general field and in the different environments, and contrast the emerging trends using the different parameterizations  presented in Sec. 3. The bin in stellar masses for each parametrization is 0.3 dex.  In each plot we slightly shifted the x-value of the medians to easily compare the errors. The bins with only one galaxy have been not considered to allow us a better interpretation of the results.

Table \ref{tab1} lists the final number of star-forming galaxies in the different environments analyzed in this work.

\begin{table}
\centering
\begin{tabular}{llcc}
  \hline
 &&  Number & \% \\
  \hline
 \multirow{3}{*}{Global environment} & GR  & 215 & $32\pm 2$\\
 & FB & 114 & $17\pm 2$\\
 & FS & 263 & $39\pm 2$\\
 & Other & 88 & $12\pm 2$\\
  \hline
  \multirow{3}{*}{Local density} &  D1 (-2$\leq$LD<-0.5) & 233 &$34\pm 2$\\
& D2  (-0.5$\leq$LD$<$0) & 214&$32\pm2$\\
  & D3  (0$\leq$LD$\leq$1.8) & 233 &$34\pm2$\\
  \hline
  \multirow{3}{*}{Halo mass} &MH1 $(11.6\leq M_{h}\leq 12.5$)  & 371 & $72\pm 2$ \\
  &MH2 ($12.5< M_{h}<13.5)$  & 109& $21\pm 2$ \\
  &MH3 (13.5$\leq M_{h}\leq 14.8$)  & 38& $7\pm 2$  \\
  \hline
  \end{tabular}
\caption{Number of galaxies above the mass and SSFR completeness limits in the different samples of PM2GC and the corresponding value in percentage. Errors on percentages are binomial. \label{tab1}}
\end{table}

\begin{figure*}
   \centering
   \includegraphics[scale=0.55]{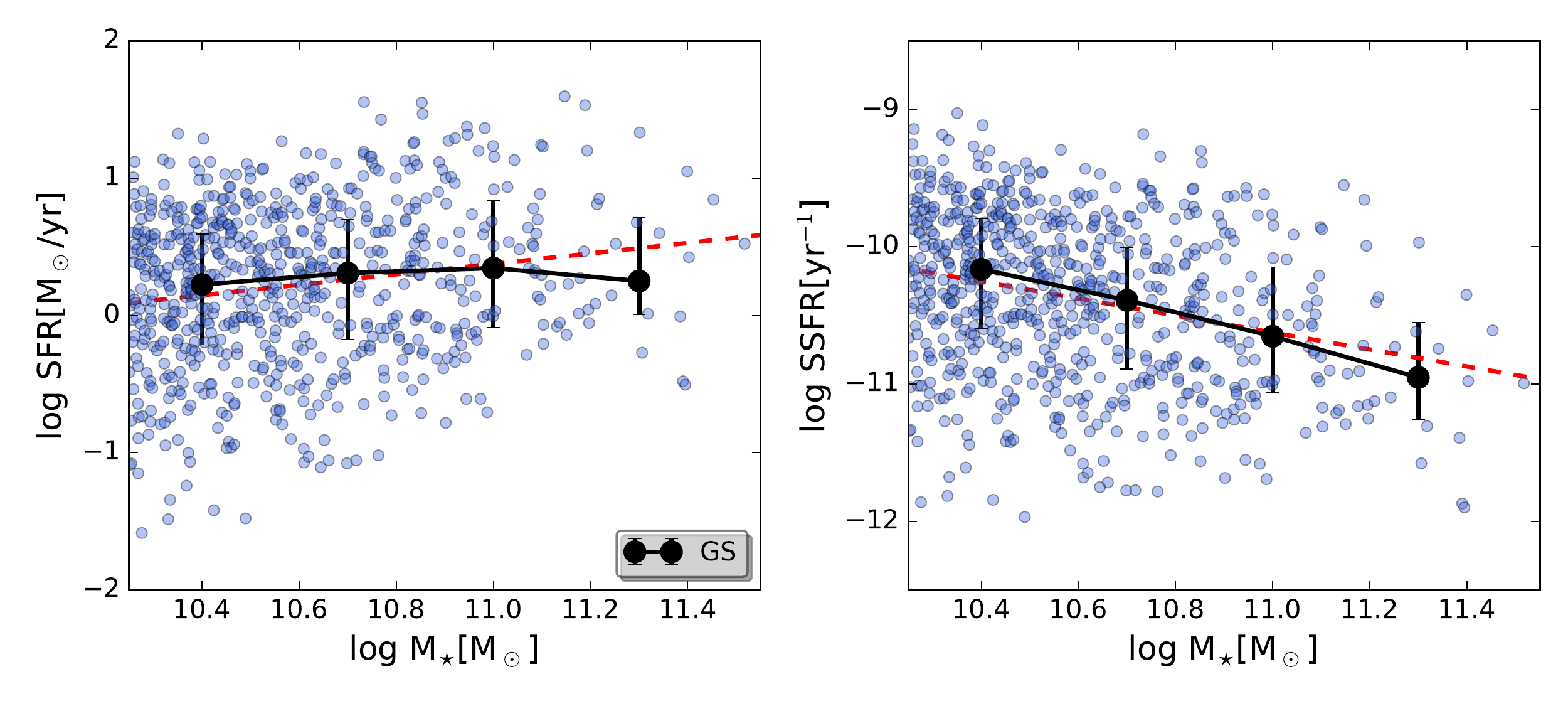}
   \caption{The SFR- (left) and SSFR- (right) $M_{\star}$ relations for galaxies above the mass and SFR limits. Blue points are the scatter plot in general sample (GS). The black big circles represent the medians in mass bins of 0.3 dex. Error bars represent the 25$^{th}$ and 75$^{th}$ percentiles of the distributions.The red dashed line correspond to the linear fit for our sample.  
   }\label{SFR_M_1}
\end{figure*}

\subsection{The SFR-$M_{\star}$ and SSFR-$M_{\star}$ relations in the different environmental parametrization}

We start our analysis by studying 
 the SFR- and the SSFR-$M_\ast$ relations for star-forming galaxies in the general sample, above our  completeness limits, shown in Figure\ref{SFR_M_1}. Error bars represent the 25$^{th}$ and 75$^{th}$ percentiles.
The relation has a quite large scatter ($\sigma$$\sim$0.6), but, in agreement with many previous determinations \citep[e.g][]{Noeske07a,Zamojski07,Zheng07,Karim11}, more massive  galaxies have systematically larger values of SFRs and lower values of SSFRs than low massive galaxies. This result shows that more massive galaxies
require more time to  double their mass than lower mass galaxies. \cite{Guglielmo15} have already explicitly compared the PM2GC SSFR-$M_\ast$ relation to that presented 
in \cite{LaraLopez13} for  an emission-line galaxy sample at $z\leq 0.1$  extracted from the Galaxy and Mass Assembly (GAMA) survey \citep{Driver11}, and the SDSS-DR7 \citep{Adelman07,Abazajian09}, showing overall good agreement.

\begin{figure*}
\centering
     \includegraphics[scale=0.6]{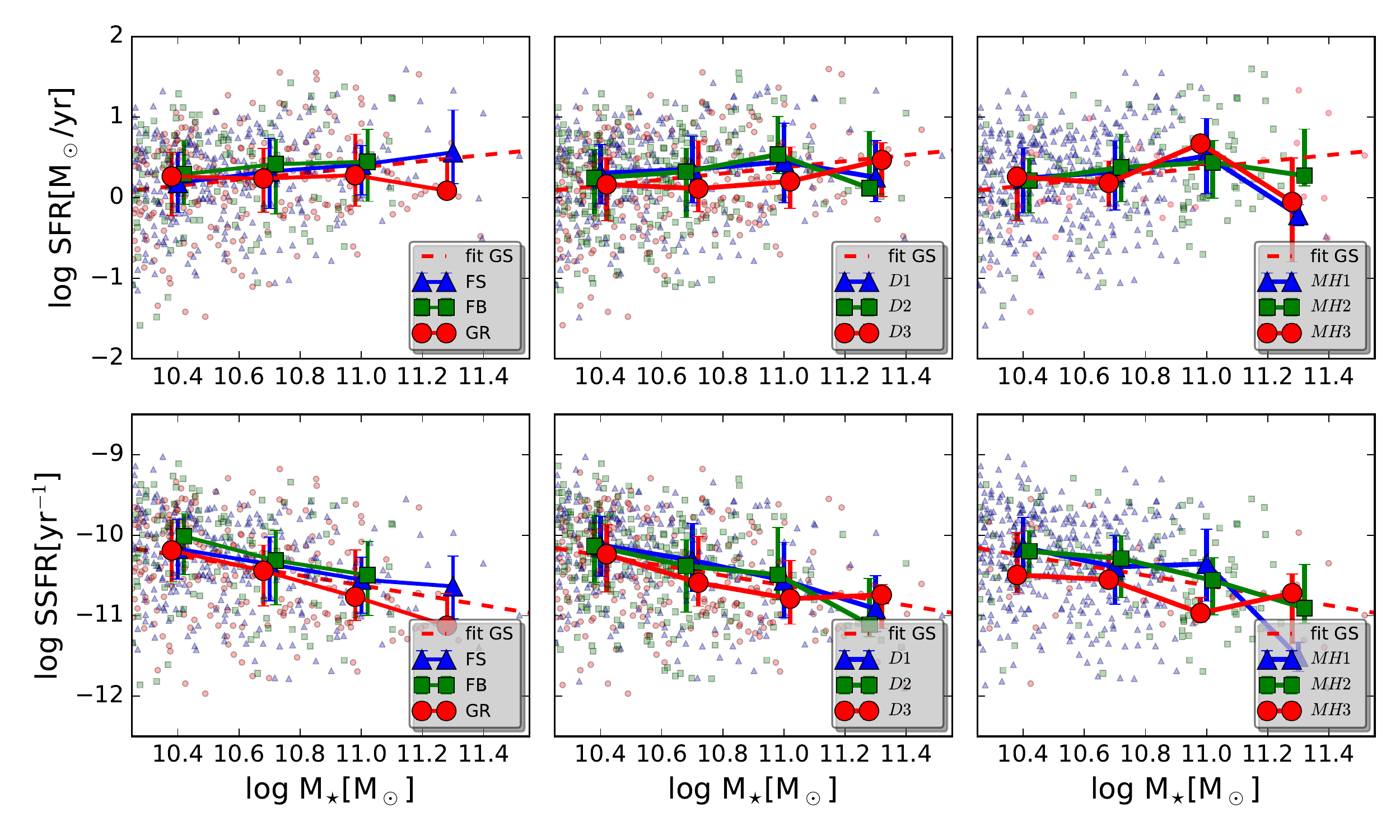}
\caption{Scatter plots and SFR -$M_{\star}$ relations (upper panels) and  SSFR -$M_{\star}$  (lower panels) in the different environmental parametrizations. Left panels: GR, FB and FS; central panels: lowest (D1), intermediate (D2) and highest (D3) bin of local densities; right panels: halo mass bins (MH1, MH2, MH3). The bin in stellar masses for each parametrization is 0.3 dex.  Error bars represent the 25$^{th}$ and 75$^{th}$ percentiles. The red dashed lines are the fit as in Figure\ref{SFR_M_1}. }\label{SFR_M_2}
\end{figure*}

We  now characterize the relation as a function of the environment.
The left upper panel of Figure  \ref{SFR_M_2} considers the global  parametrization. Galaxies in the finer environments within the general sample all share similar SFR-Mass relations. 
Trends for GR, FB and FS follow the fit obtained using the entire sample. 
Only in the most massive bin ($\log M_{\ast}/M_\odot>11$) there might be a hint that in GR the SFR declines, while in FS it increases, but  differences are only marginal ($<1\sigma$).  High ($\log M_\ast/M_\odot>11$) mass galaxies are not present in the FB sample. This result was already found and  discussed  in \citet{Calvi13}.

Similarly, also the SSFR-$M_\ast$ shown  in the lower left panel of Figure \ref{SFR_M_2},  seems not to depend on the global environment.


\begin{figure*}
   \centering
   \includegraphics[scale=0.5]{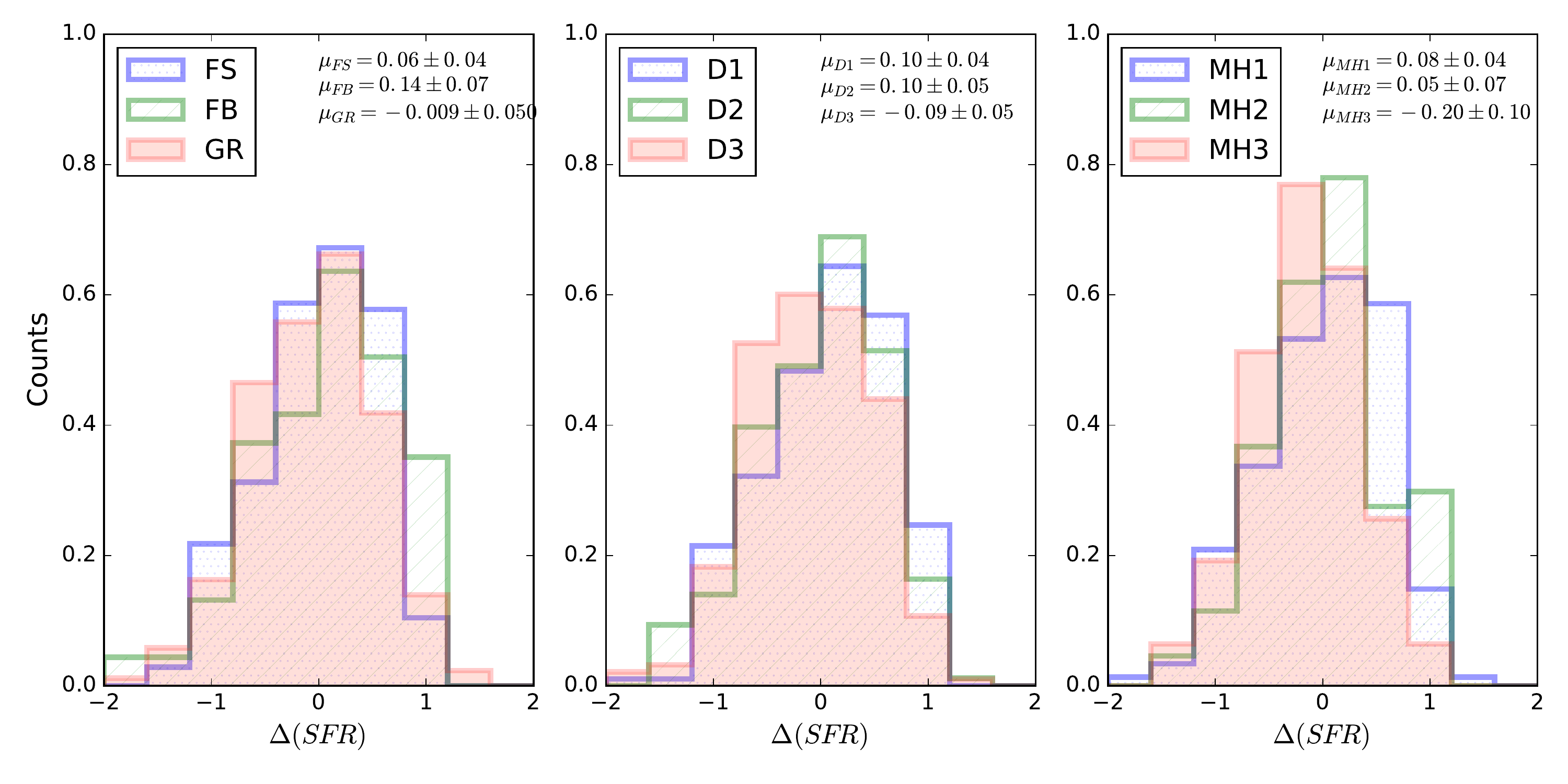}
   \caption{The normalized distributions of the differences between the galaxy SFRs and their expected values according to the fit given their mass (see text for details). In the labels median values ($\mu$) with uncertainties are reported.  Left panel: GR, FB and FS; central panel: lowest (D1), intermediate(D2) and highest (D3) bin of local densities; right panel: halo mass bins (MH1, MH2, MH3). }\label{residui}
\end{figure*}

To probe the results on a more solid statistical ground, 
we 
inspect the difference between the  SFR of each galaxy and the value derived from the fit shown as red dashed line in Figure \ref{SFR_M_1} given its mass. The  distribution of such differences for GR, FB and FS separately is shown in the left panel of Figure \ref{residui}. The median values of the three finer environments are overall in agreement: $\mu=$-0.009$\pm$0.050 for groups, $\mu=$0.06$\pm$0.04 for singles and $\mu=$0.14$\pm$0.07 for binaries. Errors on the median are calculated as $1.253\frac{\sigma}{\sqrt{N}}$ with $\sigma$ standard deviation and $N$ number of galaxies in the considered sample. The difference between GR and FB is indeed only marginal ($\sim$1$\sigma$).  Kolmogorov-Smirnov 
(K-S) tests from the pairwise comparison  of the offsets distributions are unable to find differences (P$_{GR-FS}\sim 9\%$, P$_{GR-FB}\sim 18\%$, P$_{FS-FB}\sim 37\%$).
This statistical test therefore corroborates the result that the global environment is not able to affect the SFR-M$_\ast$ relations.

The central panel of Figure \ref{SFR_M_2} shows  
 the SFR-$M_{\star}$  relation for galaxies found at different local densities. 
Subdividing the sample into 
three bins of local density, as shown in  Figure \ref{hist_ld_1} and contrasting them, median SFR values as  a function of mass seems to be in agreement within the errors. There are hints that  median values in D1 and D2 are systematically higher than in D3, but no significant differences are observed, when the errors on the median are adopted.  Similarly, also the SSFR-$M_\ast$ relation, in the lower central panel, seems not to depend on local density. In contrast, 
the distribution of the difference between the galaxy SFRs and their expected values according to the fit (central panel of Figure \ref{residui}) shows that the lowest and intermediate samples are different from the highest bin: the median value for D1 and D2 are 0.10$\pm$0.04 and  0.10$\pm$0.05, respectively, that for D3 is  -0.09$\pm$0.05. The K-S test recover significant differences only when comparing D1 and D3 (P$_{D1-D3}\sim 1\%$). In the other two cases, it is unable to spot differences (P$_{D1-D2}\sim 65\%$, P$_{D2-D3}\sim 9\%$), 

Both the statistical tests therefore evidence that the (S)SFR-M$_\ast$ relations do depend on  local density.
\begin{figure}
\centering
\includegraphics[scale=0.4]{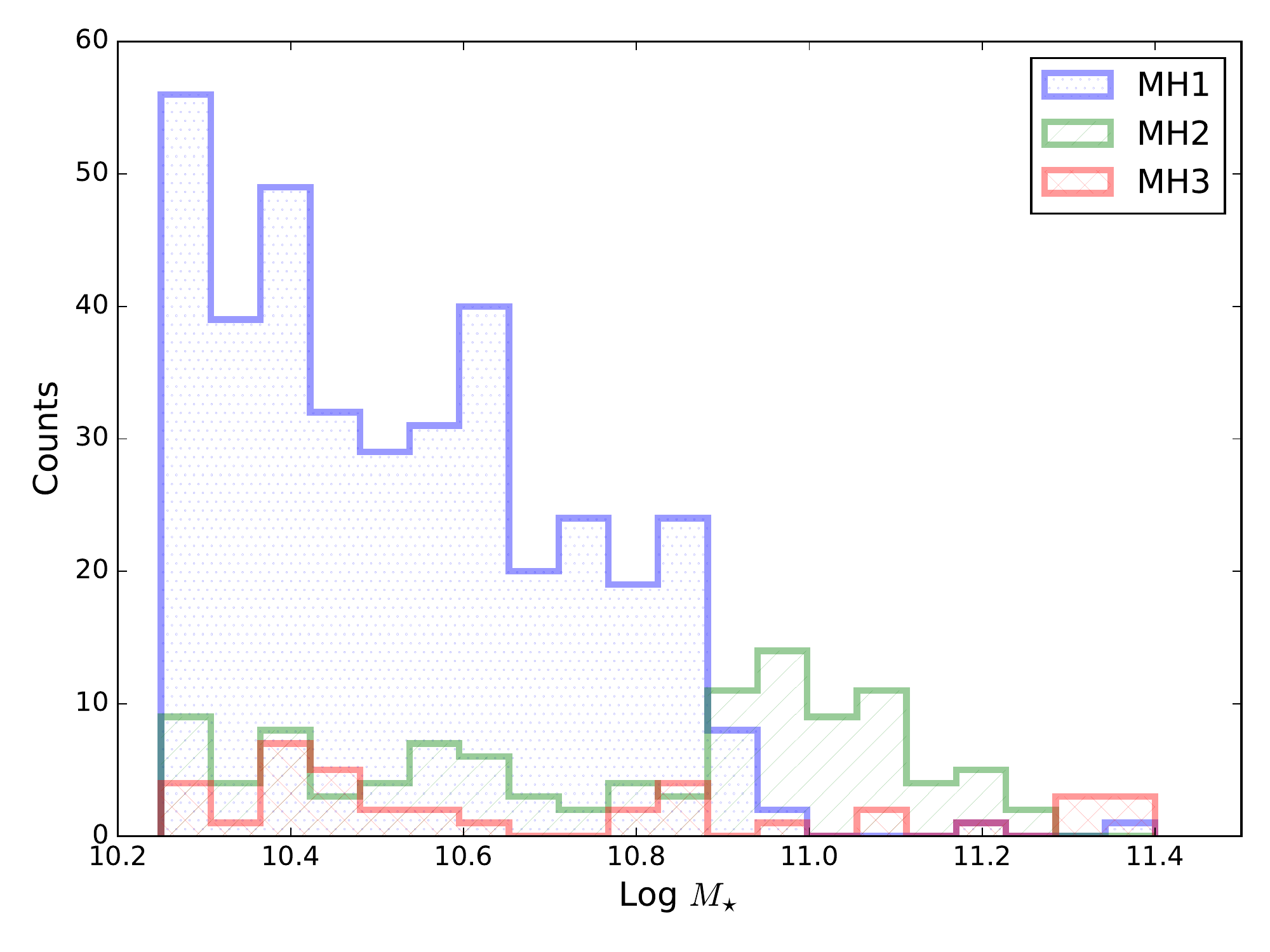}
\caption{Mass distribution of galaxies in the three different halo mass bins used in Fig. 4.
\label{hist_halo}}
\end{figure}

Finally, we consider the last parametrization we adopt.
The right panel of  Figure \ref{SFR_M_2} shows the SFR and SSFR-M$_\ast$ relations for three different bins of halo masses.

No differences are detected among the SFR-$M_\ast$ relations of galaxies in haloes of different  masses. The analysis of the residuals (right panel of Figure\ref{residui}) supports the results when we compare MH1 and MH2 ($\mu=$ 0.08$\pm$0.04 for MH1 and 0.05$\pm$0.07 for MH2). In contrast, in MH3 the median is significantly lower (-0.20$\pm$0.1). 
Similar results are obtained for the SSFR-$M_\ast$ relation (bottom panel of Fig. 4):  at any given mass galaxies in MH3 have systematically lower SSFR values than galaxies in the other bins. 
Note however the poor number statistic of MH3, which includes only 38 galaxies. 
It is also interesting to inspect the mass distribution in the different halo mass bins. Indeed, the mass distribution of these samples has never been inspected (for the mass distribution of galaxies in the different local and global environments see \citealt{Vulcani12, Calvi13}). Figure \ref{hist_halo} shows  the distributions of stellar masses in the three bin of halo mass. In MH1 98\% of galaxies have $M_\ast< 10^{10.9} M_{\odot}$, while in MH2 50\% of  galaxies have $M_{\ast}>10^{10.9}M_{\odot}$. MH2 actually consists almost half of the galaxy population of MMGs with high stellar masses. 
In MH3, which actually is poorly populated, the fraction of galaxies with $M_{\ast}>10^{10.9}M_{\odot}$ is $\sim 29\%$. 

The SFR seems to be independent on halo mass but more related to the different mass distribution of galaxy population in the different haloes. As the SSFR is the ratio between the SFR and mass, this explains the systematically lower SSFR-$M_{\star}$ relation in MH3.
Performing the analysis of the residuals (right panel of Figure 5), the KS test detects differences only when MH1 and MH3 are compared. 

To conclude, in this section we have found that the different parametrizations of environment have different impact on the SFR and SSFR-Mass relations: considering the global environment, no differences are detected, considering the halo mass, the different mix of galaxies with different masses have an incidence on SFRs and SSFRs. Finally, when considering the local density, statistically robust differences emerge between D1 and D3 for both relations.  

In the next section, before discussing this result, we will inspect whether these relations also depend on an intrinsic property of the galaxies, that is galaxy morphology.

\subsection{The morphological dependencies}

\begin{table}
 \centering
\begin{adjustbox}{center, width=\columnwidth-10pt}
\begin{tabular}{llccc}
\hline
 &&\multicolumn{3}{c}{\% morphological types}  \\
&  & Ellipticals & S0s & Late-types \\
\hline
 \multirow{4}{*}{Global environment} & GR   &  $19\pm 3$ & $23\pm 3$ & $58\pm 4 $\\
 & FB &  $15\pm 4$ & $15\pm 4$& $70\pm 5$\\
 & FS & $13\pm 2 $ & $21\pm 3$ & $66\pm 3$\\
& All GS & $16\pm 1 $& $21\pm 2$ & $63\pm 2 $\\
  \hline
  \multirow{3}{*}{Local density} &  D1  & $15\pm 2 $ & $17\pm 3$&$68\pm 3$ \\
  & D2  & $18\pm 2$ &$19\pm 2$ &$63\pm 3 $\\
  & D3   & $16\pm 3$ & $27\pm 3$ &$57\pm 3 $\\
  \hline
  \multirow{3}{*}{Halo mass} &MH1 & $14\pm 2$& $20\pm 2$& $66\pm 6 $  \\
  &MH2  & $18\pm 2$ & $25\pm 2$ &$57\pm 5$ \\
  &MH3  & $ 21\pm 2$ & $24\pm 2$ &$55\pm 5 $ \\
  \hline
\end{tabular}
\end{adjustbox}
\caption{Fractions of galaxies above the mass and SSFR limits of each morphological type in the different samples.  Errors are binomial.}\label{tab2}
\end{table}

To investigate the connection between morphology and star formation, we now investigate the  morphological distribution of the star forming galaxies in the different environments and if the incidence of the morphological type might drive the SFR- and SSFR-mass relations.

\subsubsection{Morphological fractions in different environments}\label{frac}

\begin{figure*}
\centering
 \includegraphics[scale=0.55]{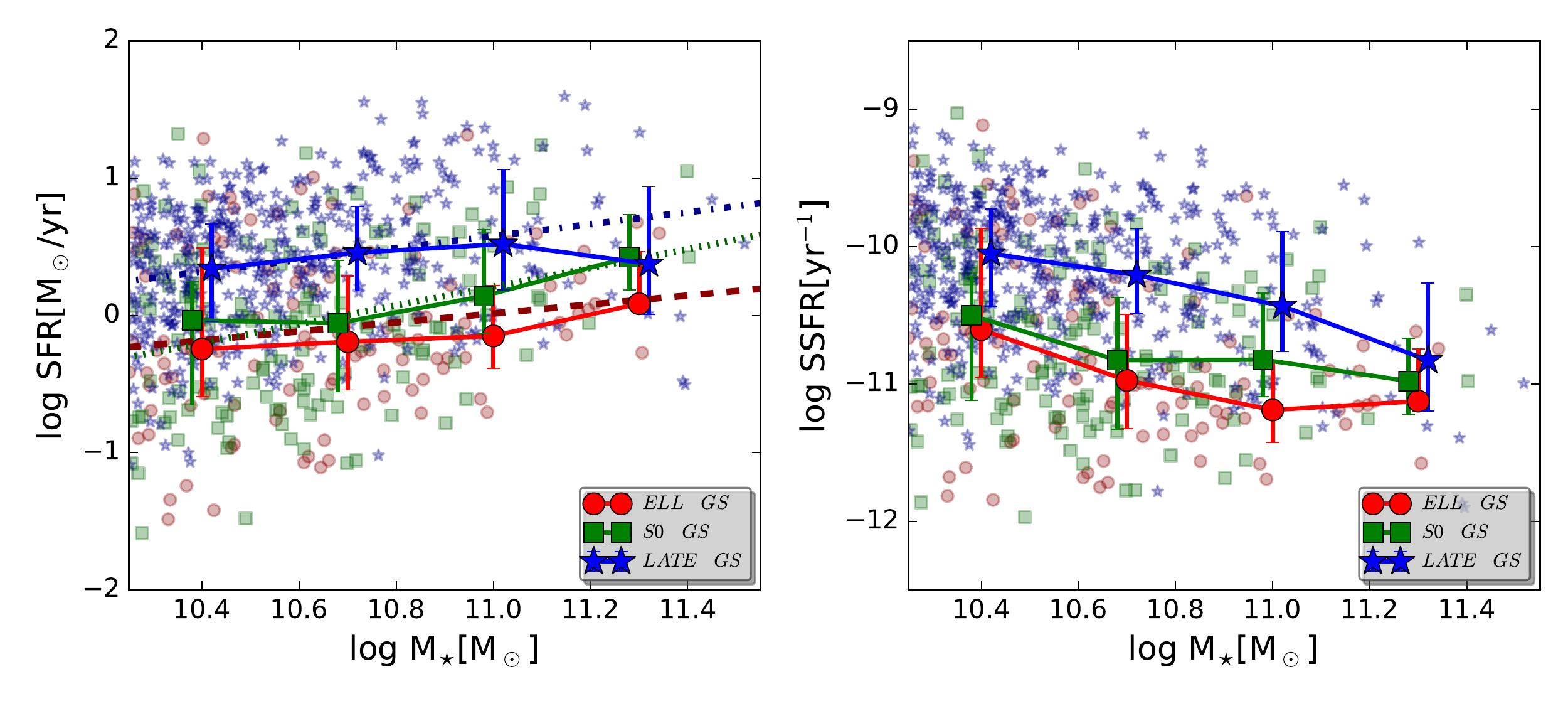}
 \caption{Scatter plots and SFR- and SSFR-mass relations for galaxies of different morphological types in the GS. Blue stars are late-types, green squares S0s and red circles ellipticals. Error bars are the $25^{th}$ and $75^{th}$ percentiles. For each morphological type we also show the linear fits of the SFR-$M_{\ast}$ relations (the dashed-dotted blue line for late-types, the dotted green line for S0s and the dashed red line for ellipticals).}\label{SFR_M_5}
\end{figure*}

 Table \ref{tab2} lists the fraction of each morphological type among star-forming galaxies above the mass and SFR completeness limits. In the GS,  ellipticals represent 16$\pm$1\% of all star-forming galaxies. Their fraction weakly depends on the global environment: it progressively increases going from the least to the most massive environments, i.e going from FS galaxies (13$\pm$2\%) to FB (15$\pm$4\%), to GR (19$\pm$3\%).  In contrast, ellipticals are found in similar proportions in the three bins of local densities ($\sim$15-18\%). When considering the halo mass  parametrization the fraction of ellipticals in lower bin is low but not negligible, as we could expect in halo masses below $\rm M_{h}\leq 10^{12.5} h^{-1} M_{\odot}$, and they represent the 14\% of all star forming galaxies. In contrast,  at intermediate and higher halo masses they are among 18-21\% of the sample. We caution the reader that this result could be a stellar mass effect and not an environmental effect because, as we calculated the halo masses basing on the stellar masses, is very difficult differentiating the relative effects of mass and environment. In any case, the non negligible presence of star-forming ellipticals was already found in \cite{Vulcani15} and suggests that for these galaxies morphology changes before star formation is completely  shut off. \cite{Vulcani15} found that this morphological transition is mainly due to the fading and total or partial removal of the disk. According to their SFH, these star forming galaxies descend from star forming late-type galaxies and not from rejuvenated passive early-types. S0s represent 21$\pm$2\% of all star forming galaxies in the GS. Their percentage is lower among FB (15$\pm$4\%) than among GR and FS (21-23\%).  Their incidence also depends on local density: it ranges from 17$\pm$3\% in D1 and $19\pm 2$ in D2 to 27$\pm3$\% in D3. Considering the halo masses,   in MH1 the fraction of S0 is 20\% while in MH2 and MH3 agrees within the errors (25\% and 24\%). 
The fraction of early types therefore turns out to be almost constant in MH2 and MH3 ($\sim 45\%$), but lower in MH1 ($\sim 34\%$). We checked that 50\% of the galaxies in MH2 are massive MMGs, and also MH3 contains a good percentage of these galaxies. The presence of a higher fraction of star-forming galaxies with an early type morphology is therefore a hint that the formation of central spheroid, and thus the morphological transformation, is halo mass-dependent and that in massive halos this process is more efficient than the quenching of star formation. This result agrees with studies about the bulge and disc colours in cluster galaxies which suggest a scenario in which the formation of bulges is regulated by internal processes, while the discs are much more affected by the environment \citep{Hudson10,Head14}.

Focusing on late-type galaxies, we find that they dominate in number in all the environments. This result was expected given the tight correlation between star-formation rate and morphology (e.g. \citealt{RobertsHaynes94,Blanton03}).  
In the GS, they are 63$\pm2$\% of the total. 
Their fraction is higher in FB and FS (70$\pm$5\% and 66$\pm$3\%, respectively) than in GR (58$\pm$4\%). In interacting binary galaxies we are expecting the action of tidal stripping and tidal friction. Thus, a larger number of late-type galaxies could be a consequence of the development of bars and/or spiral arms because of the diffuse tidal gas ejected during collisions.                                          
Their fraction also depends on  local density (68$\pm$3\% in D1, 63$\pm$3 in D2 and lower, 57$\pm$3\%, in D3) but, within the errors,  only marginally on halo mas, ranging from 55 to 66\%. 

 \subsubsection{The SFR-$M_{\star}$ and SSFR-$M_{\star}$ relations for the different Hubble types in different environments }\label{sec:sfr_m_mor}
\subsubsection {Fixed environment}



Figure \ref{SFR_M_5} shows the SFR- and SSFR- $M_\ast$ relations for  ellipticals, S0s and late-type galaxies in the GS. 
Galaxies of the different types populate different regions of the  SFR-$M_{\star}$ plane.  
At any given mass, late-type galaxies have systematically the highest SFR, followed by S0s and ellipticals. 
Indeed,  when we consider the median values  as a function of mass for the  three galaxy populations, we find that at any given mass, median SFR values increase from ellipticals, to S0s, to late-type galaxies. Correlations overlap within the uncertainties (which represent the $25^{th}$ and $75^{th}$ percentiles), but trends are clear.  

Performing a least square linear fit, we find that  ellipticals have the flattest relation {\bf ($\log SFR=(0.32\pm0.07) \times \log M_\ast/M_\odot-(3.58\pm0.73)$)}, followed by late types {\bf $\log SFR=( 0.43\pm0.02) \times \log M_\ast/M_\odot-(4.21\pm0.18)$)}and S0s {\bf ($\log SFR=(0.68\pm0.06) \times \log M_\ast/M_\odot-(7.32\pm0.68)$)}.

The shift observed among the different morphological types generates the large spread in the total SFR-mass relation observed in Figure 3.

Considering the SSFR, we find that  the relation for late-type galaxies is systematically higher than that for ellipticals and S0s, showing how late-types are much more efficient in forming stars than early-type galaxies. 

In both cases, S0s occupy the region between ellipticals and late-types, even if  the trends are fairly close to those of ellipticals. 

\begin{figure*}
\centering
\includegraphics[scale=0.6]{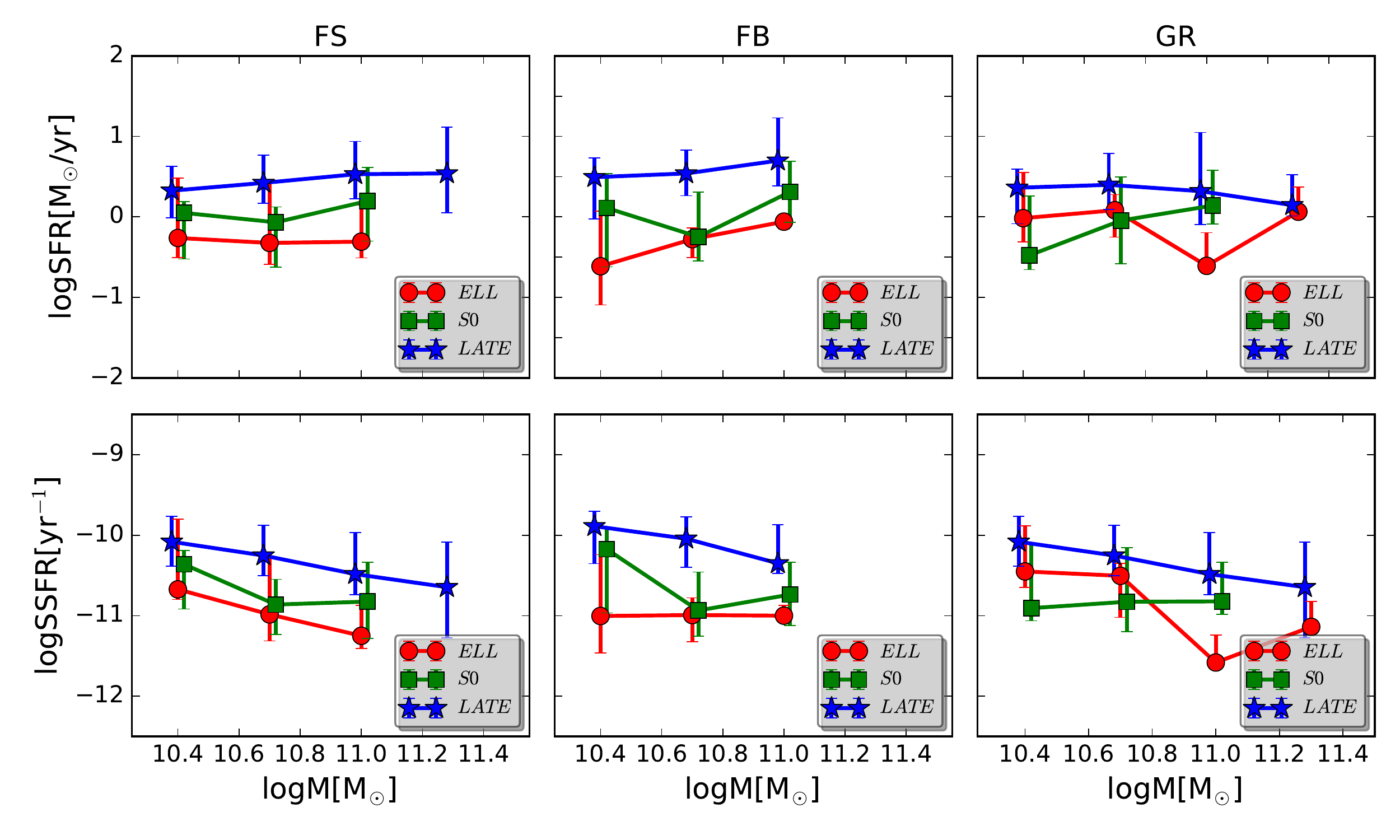}
\caption{SFR-$M_{\star}$ (upper panels) and SSFR- $M_{\star}$ (lower panels) fixing the morphological type for galaxies in GR (right panels), FB (central panels) and FS (left panels). Points are the medians in the mass bins: red circles are ellipticals, green squares are S0s, blue stars are late-types.  Error bars represent the 25$^{th}$ and 75$^{th}$ percentiles. \label{SFR_M_7}}
\end{figure*}

These findings hold also when considering galaxies in the finer environments, as shown in 
Figure \ref{SFR_M_7}. In FS, FB and GR separately, at any given mass both the median SFR and SSFR values increase from ellipticals, to S0s, to late-types. In FB, the difference between late-type and early type is the largest. Similar results are obtained when we consider local densities and halo masses (plots shown in Appendix \ref{appendix}): in any given environment late type galaxies are characterized by the highest values of SFR and S0s are always in between late-types and ellipticals. 

\subsubsection{Fixed morphological type}

\begin{figure*}
\centering
\includegraphics[scale=0.65]{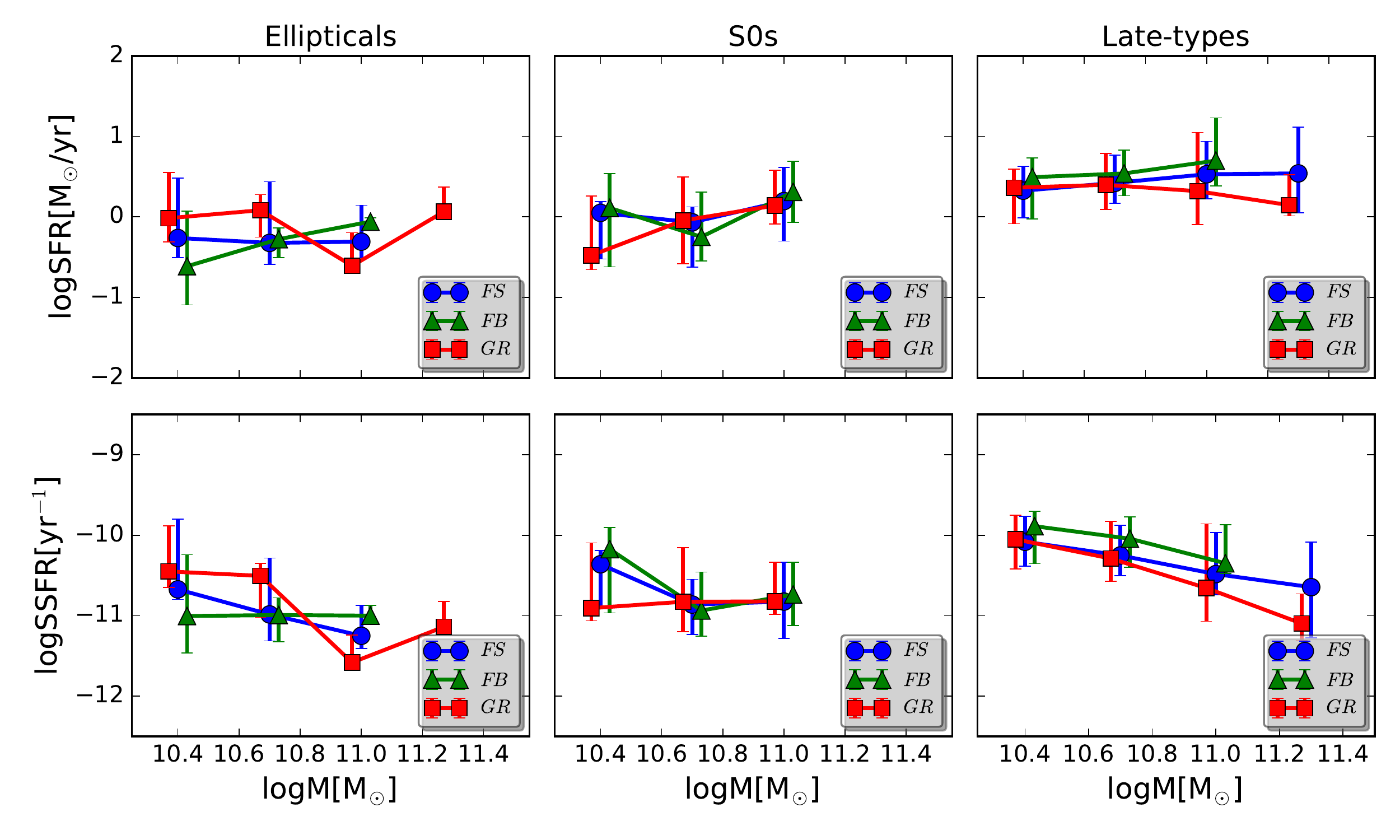}
\caption{SFR-$M_{\star}$ (upper panels) and SSFR- $M_{\star}$ (lower panels) for ellipticals (left panels), S0s (central panels) and late-types (right panels) galaxies fixing the environment: GR (square points), FB (triangle points) and FS (circle points). Points are medians in the mass bins. Error bars represent the 25$^{th}$ and 75$^{th}$ percentiles. \label{SFR_M_8}}
\end{figure*}

\begin{figure*}
\centering
\includegraphics[scale=0.65]{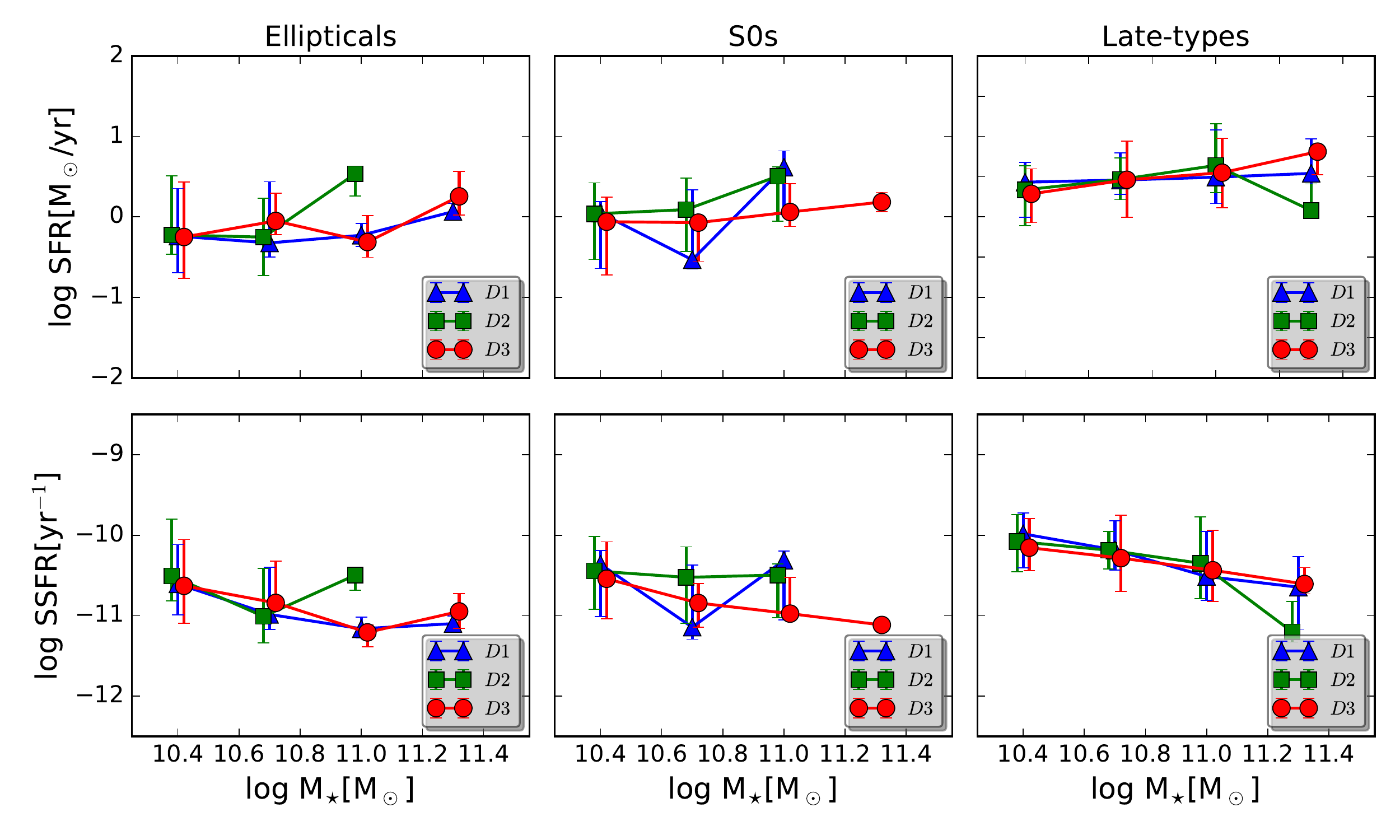}
\caption{Comparison between the SFR- and SSFR- $M_{\star}$ in the three bin of local density, D1 ($-2\leq LD\leq -0.5$), D2 ($-0.5 < LD < 0$) and D3 ($0\leq LD\leq 1.8$) fixing the morphological type. In each panel the blue triangle points are the medians in the bin D1, the green squares points the medians in D2 and red circle points the ones in D3. Error bars represent the 25$^{th}$ and 75$^{th}$ percentiles. 
}\label{SFR_M_9}
\end{figure*}

Figure \ref{SFR_M_8} shows the trends for galaxies of each morphology separately, in the different global environments. Overall, given the morphology, there is no clear trends with global environment and very little variations are observed across them.  Among late types,  massive galaxies in groups might have lower median SFRs and SSFRs than  single galaxies. Similarly, when considering the local density parametrization (Figure \ref{SFR_M_9}), the values of SFR and SSFR for the three morphological types are essentially comparable in each local density bin. 
The differences detected between D1 and D3 when considering all galaxies together (central panel of Figure \ref{SFR_M_1}) and the lack of a clear trend support the fact that  any physical environmental mechanism that regulates the star-formation activity is not the main driver in affecting the SFR for a galaxy of a given mass and morphology.

Similar results are also obtained when we look at the halo mass parametrization as discussed in Appendix \ref{appendix}.

\begin{figure*}
\centering
 \includegraphics[scale=0.6]{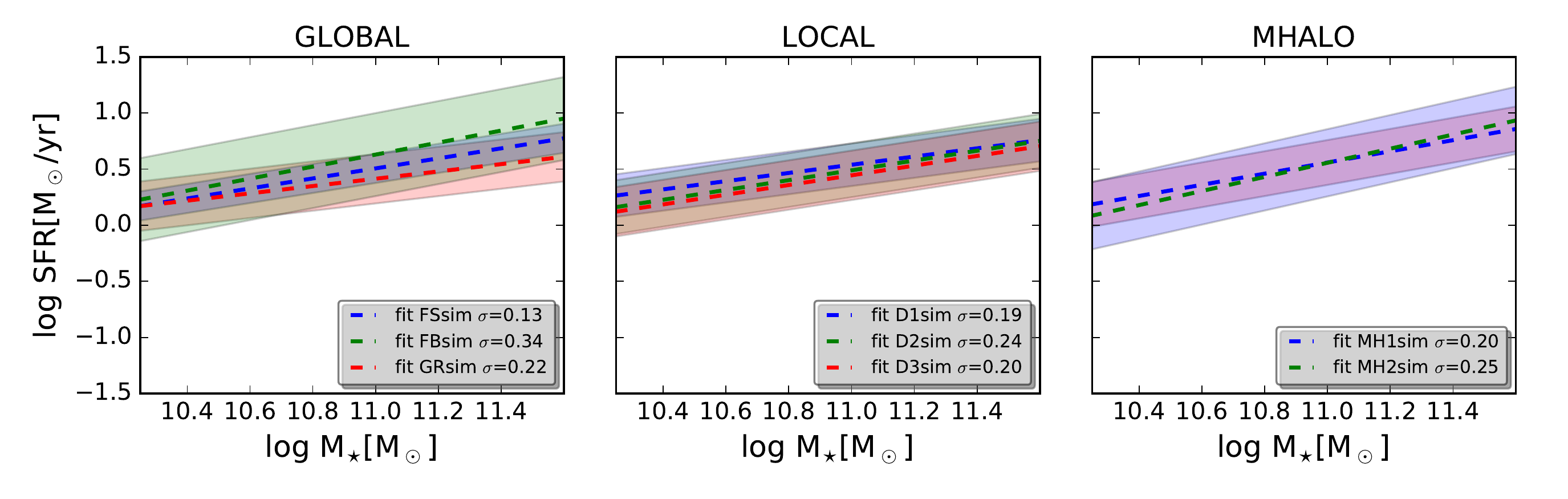}
 \caption{Comparison between the linear fits of the simulated SFR-mass relations. The simulated fits are obtained considering the mean slopes and the mean intercepts on 100 random extractions for each samples. The samples of galaxies, in each environmental parametrization, have the same morphological proportions: 20\% of ellipticals, 20\% S0s and 60\% late-types. In the legend are also indicated the standard errors of each fit. }\label{SFR_M_new}
\end{figure*}
Therefore, the environment seems not to affect SFR-$M_{\star}$ and SSFR-$M_{\star}$ relations of each morphological type separately, but it does influence the morphological fractions, as seen in Sec. \ref{frac}.  
To disentangle the two effects, 
we investigate the SFRs extracting sample of galaxies, with the same  morphological proportions, in each environmental parametrization. We choose 20\% elliptical galaxies, 20\% S0 galaxies and 60\% late-type galaxies. These fractions are overall  representative of the make up of galaxies in our samples (in each samples  the late-types are always the dominant population).
We performed 100 random extractions and we show in Figure \ref{SFR_M_new} the comparison between the simulated fits on these extractions (the slopes and intercepts are the mean values of the 100 fits) for each environmental parametrization. Note that, given the low number statistics in MH3 (38 galaxies) the random extraction is meaningless and we do not plot the results.


The trends of SFR-M$_{\star}$ relations does not change within the errors fixing the morphological mix in each environmental parametrization. Note that choosing a different morphological mix (e.g. 40\% ellipticals, 40\% S0s and 20\% late-types) results stay robust. 

We can therefore conclude that the morphological content among different environments might be responsible of the offsets observed in Figure \ref{SFR_M_2}.


\section{Summary and conclusions}
In this paper we have analyzed the SFR- and SSFR- mass relations for star-forming galaxies drawn from the PM2GC \citep{Calvi12} catalog (0.038 $\lesssim$z$\lesssim$0.104), considering three different parameterizations of environment (global environment, local density and halo mass). We have also investigated how these relations depend on the galaxy morphological type. Our aim was to understand to what extent the physical processes that regulate the star formation are due to environmental influences and/or morphological signatures. We have included all galaxies above the mass completeness limit of log$(M_{\ast}/M_{\odot})\geq$10.25 and SSFR  threshold of 10$^{-12} yr^{-1}$. The main results of this work can be summarized as follows:
\begin{itemize}
\item The SFR-$M_{\star}$ and SSFR-$M_{\star}$ relations for star-forming galaxies do not depend on the environment, when the global parametrization is used. When the halo mass is considered, galaxies in massive haloes tend to be systematically less star forming, at any given mass. Trends in the (S)SFR-$M_{\star}$ relations might reflect the switch from MMGs with high stellar mass to low-mass satellites. In contrast, using the local density parametrization, we do detect different trends for galaxies at different densities: at any given stellar mass, objects in low density environments are characterized by  systematically higher values of SFR than objects in denser environments. Our results therefore  agree with e.g. \cite{VonderLinden10}, 
but are at odds with, e.g., \cite{Peng10}. Note however that the different selection criteria adopted by the different authors to assemble the samples prevent us from fairly compare results.
Our point-distributions of galaxies show a wide spread. Simulations by e.g. \cite{Dutton10, Sparre15,RodriguezPuebla16} showed that
the scatter in the SFR sequence reflects variations in the gas accretion history.
Our result suggests the coexistence of different populations, in agreement to, e.g., \citet{Oemler16}. 
In particular, we have found that the wide spread in the SFR-$M_{\star}$ and SSFR-$M_{\star}$ relations can be explained in terms of morphological type:  the relation for late-type galaxies is systematically higher than that of S0s and ellipticals. 

\item The morphological fractions of star forming galaxies depend on environment. 
\begin{itemize}
\item A non negligible fraction (16$\pm$1\%) of galaxies have elliptical morphology.  Their incidence increases with global environments, from FS ($\sim$13\%) to GR ($\sim$19\%), and changes with halo mass, going from 14$\pm$2\% in the lowest bin to 21$\pm$2\% in the highest one.
In contrast, it does not depend on local density ($\sim$16\%), within the errors.

\item  The fraction of S0s is suppressed in FB ($\sim$15\%), and very similar in the other global environments ($\sim 22\%$). The same fraction increases with local density (from $\sim17\%$ to $\sim 27\%$). Considering halo masses, 
S0 are more common in haloes of intermediate mass  than in the other bins ($\sim24\%$ in MH3 and $\sim 20\%$ in MH1).  This result might be driven by the fact that there are no massive binaries and massive galaxies are generally early-types \citep[see, e.g.,][]{Vulcani15}.

\item
Late type galaxies dominate in number in all the environments. This result was expected given the tight correlation between star-formation rate and morphology (e.g. \citealt{RobertsHaynes94,Blanton03}).  The fraction of late-types is lower in GR and at high densities ($\sim 58\%$) than in the other environments of the same parametrization. 
The fraction of late types is maximum in FB. In interacting binary galaxies we are expecting the action of tidal stripping and tidal friction. Thus, a larger number of late-type galaxies could be a consequence of the development of bars and/or spiral arms because of the diffuse tidal gas ejected during collisions.
Within the errors, the fraction of late types is constant across the different halo masses. 
\end{itemize}

\item The SFR-$M_{\star}$ and SSFR-$M_{\star}$ relations of each morphological type do not change with environment, but we detected significant differences among the relations for the different morphologies. Furthermore, we also found that the small difference in the different parametrizations are just the result of their different morphological proportions.

\end{itemize}

Our results therefore point to a scenario in which morphology rather than environment drives the correlation between the star forming properties and stellar mass and its spread. Recently, also \citet{Bait17} argue that the morphological T-type is a strong indicator of SSFR, with a weak dependence on the environment.

The results presented in this paper show that, 
 up to the group scale, 
 when the environmental definition adopted is somehow related to the halo mass (either using observational or theoretical approaches)
no strong environmental effect is able to influence such correlations, suggesting that the different processes acting on galaxies in the different environments leave similar signatures on the star forming properties of galaxies. The same processes, however, do suppress star formation in denser environments, as the morphological mix strongly depend on environment. 
These processes have  to act on similar time scales on star forming properties and morphology, given that galaxies of the same Hubble type have similar star formation values in all the environments. Note that, instead,  \cite{Paccagnella16} have shown that, at similar redshifts,  the SFR-mass relation in clusters is systematically shifted toward lower values compared to the field. Taken together, these results suggest that only cluster specific processes, such as strangulation, are able to halt the star formation on long time scales (2-5 Gyr), while any other process acting on general field galaxies does not leave significant evidences. 

In contrast, when a environmental parametrization based on the local density is adopted, our data unveil a dependence of the SFR-mass relation on environment. Local density does not track well the global environment of the galaxies and our findings point to a scenario where processes taking place on local scales, such as galaxy-galaxy interaction or tidal effects, is more important than the global environment in shaping the properties of the galaxies, in agreement with previous results (e.g. \citealt{Vulcani12,Valentinuzzi11,Poggianti09,Balogh04,Martinez08}).

\section{Acknowledgments}

We thank Joe Liske, Simon Driver and the whole MGC team
for making easily accessible a great dataset. We also thanks Angela Paccagnella for providing us with the halo mass estimates of the PM2GC. R.C. thanks Jos\'e Miguel Rodr\'iiguez Espinosa to have made possible making this work. In addition, R.C. acknowledges the support from the Instituto de Astrof{\'{\i}}sica de Canarias (project AYA2015-70498-C2-1-R).
B.V. acknowledges the support from an Australian Research Council Discovery Early Career Researcher Award (PD0028506).

\bibliographystyle{mnras}
\bibliography{bibrosa2}

\begin{thebibliography}{}
\makeatletter
\relax
\def\mn@urlcharsother{\let\do\@makeother \do\$\do\&\do\#\do\^\do\_\do\%\do\~}
\def\mn@doi{\begingroup\mn@urlcharsother \@ifnextchar [ {\mn@doi@}
  {\mn@doi@[]}}
\def\mn@doi@[#1]#2{\def\@tempa{#1}\ifx\@tempa\@empty \href
  {http://dx.doi.org/#2} {doi:#2}\else \href {http://dx.doi.org/#2} {#1}\fi
  \endgroup}
\def\mn@eprint#1#2{\mn@eprint@#1:#2::\@nil}
\def\mn@eprint@arXiv#1{\href {http://arxiv.org/abs/#1} {{\tt arXiv:#1}}}
\def\mn@eprint@dblp#1{\href {http://dblp.uni-trier.de/rec/bibtex/#1.xml}
  {dblp:#1}}
\def\mn@eprint@#1:#2:#3:#4\@nil{\def\@tempa {#1}\def\@tempb {#2}\def\@tempc
  {#3}\ifx \@tempc \@empty \let \@tempc \@tempb \let \@tempb \@tempa \fi \ifx
  \@tempb \@empty \def\@tempb {arXiv}\fi \@ifundefined
  {mn@eprint@\@tempb}{\@tempb:\@tempc}{\expandafter \expandafter \csname
  mn@eprint@\@tempb\endcsname \expandafter{\@tempc}}}

\bibitem[\protect\citeauthoryear{{Abazajian} et~al.,}{{Abazajian}
  et~al.}{2009}]{Abazajian09}
{Abazajian} K.~N.,  et~al., 2009, \mn@doi [\apjs]
  {10.1088/0067-0049/182/2/543}, \href
  {http://adsabs.harvard.edu/abs/2009ApJS..182..543A} {182, 543}

\bibitem[\protect\citeauthoryear{{Abramson}, {Kelson}, {Dressler}, {Poggianti},
  {Gladders}, {Oemler}  \& {Vulcani}}{{Abramson} et~al.}{2014}]{Abramson14}
{Abramson} L.~E.,  {Kelson} D.~D.,  {Dressler} A.,  {Poggianti} B.,  {Gladders}
  M.~D.,  {Oemler} Jr. A.,   {Vulcani} B.,  2014, \mn@doi [\apjl]
  {10.1088/2041-8205/785/2/L36}, \href
  {http://adsabs.harvard.edu/abs/2014ApJ...785L..36A} {785, L36}

\bibitem[\protect\citeauthoryear{{Adelman-McCarthy} et~al.,}{{Adelman-McCarthy}
  et~al.}{2007}]{Adelman07}
{Adelman-McCarthy} J.~K.,  et~al., 2007, \mn@doi [\apjs] {10.1086/518864},
  \href {http://adsabs.harvard.edu/abs/2007ApJS..172..634A} {172, 634}

\bibitem[\protect\citeauthoryear{{Allen}, {Kacprzak}, {Glazebrook}, {Tran},
  {Spitler}, {Straatman}, {Cowley}  \& {Nanayakkara}}{{Allen}
  et~al.}{2016}]{Allen16}
{Allen} R.~J.,  {Kacprzak} G.~G.,  {Glazebrook} K.,  {Tran} K.-V.~H.,
  {Spitler} L.~R.,  {Straatman} C.~M.~S.,  {Cowley} M.,   {Nanayakkara} T.,
  2016, \mn@doi [\apj] {10.3847/0004-637X/826/1/60}, \href
  {http://adsabs.harvard.edu/abs/2016ApJ...826...60A} {826, 60}

\bibitem[\protect\citeauthoryear{{Aragon-Salamanca}, {Ellis}, {Couch}  \&
  {Carter}}{{Aragon-Salamanca} et~al.}{1993}]{Aragon93}
{Aragon-Salamanca} A.,  {Ellis} R.~S.,  {Couch} W.~J.,   {Carter} D.,  1993,
  \mn@doi [\mnras] {10.1093/mnras/262.3.764}, \href
  {http://adsabs.harvard.edu/abs/1993MNRAS.262..764A} {262, 764}

\bibitem[\protect\citeauthoryear{{Bait}, {Barway}  \& {Wadadekar}}{{Bait}
  et~al.}{2017}]{Bait17}
{Bait} O.,  {Barway} S.,   {Wadadekar} Y.,  2017, preprint, \href
  {http://adsabs.harvard.edu/abs/2017arXiv170700568B} {} (\mn@eprint {arXiv}
  {1707.00568})

\bibitem[\protect\citeauthoryear{{Baldry}, {Glazebrook}, {Brinkmann},
  {Ivezi{\'c}}, {Lupton}, {Nichol}  \& {Szalay}}{{Baldry}
  et~al.}{2004}]{Baldry04}
{Baldry} I.~K.,  {Glazebrook} K.,  {Brinkmann} J.,  {Ivezi{\'c}} {\v Z}.,
  {Lupton} R.~H.,  {Nichol} R.~C.,   {Szalay} A.~S.,  2004, \mn@doi [\apj]
  {10.1086/380092}, \href {http://cdsads.u-strasbg.fr/abs/2004ApJ...600..681B}
  {600, 681}

\bibitem[\protect\citeauthoryear{{Balogh}, {Schade}, {Morris}, {Yee},
  {Carlberg}  \& {Ellingson}}{{Balogh} et~al.}{1998}]{Balogh98}
{Balogh} M.~L.,  {Schade} D.,  {Morris} S.~L.,  {Yee} H.~K.~C.,  {Carlberg}
  R.~G.,   {Ellingson} E.,  1998, \mn@doi [\apjl] {10.1086/311576}, \href
  {http://adsabs.harvard.edu/abs/1998ApJ...504L..75B} {504, L75}

\bibitem[\protect\citeauthoryear{{Balogh}, {Baldry}, {Nichol}, {Miller},
  {Bower}  \& {Glazebrook}}{{Balogh} et~al.}{2004}]{Balogh04}
{Balogh} M.~L.,  {Baldry} I.~K.,  {Nichol} R.,  {Miller} C.,  {Bower} R.,
  {Glazebrook} K.,  2004, \mn@doi [\apjl] {10.1086/426079}, \href
  {http://cdsads.u-strasbg.fr/abs/2004ApJ...615L.101B} {615, L101}

\bibitem[\protect\citeauthoryear{{Bamford} et~al.,}{{Bamford}
  et~al.}{2009}]{Bamford09}
{Bamford} S.~P.,  et~al., 2009, \mn@doi [\mnras]
  {10.1111/j.1365-2966.2008.14252.x}, \href
  {http://adsabs.harvard.edu/abs/2009MNRAS.393.1324B} {393, 1324}

\bibitem[\protect\citeauthoryear{{Bell} \& {de Jong}}{{Bell} \& {de
  Jong}}{2001}]{BelldeJong01}
{Bell} E.~F.,  {de Jong} R.~S.,  2001, \mn@doi [\apj] {10.1086/319728}, \href
  {http://adsabs.harvard.edu/abs/2001ApJ...550..212B} {550, 212}

\bibitem[\protect\citeauthoryear{{Berlind} et~al.,}{{Berlind}
  et~al.}{2006}]{Berlind06}
{Berlind} A.~A.,  et~al., 2006, \mn@doi [\apjs] {10.1086/508170}, \href
  {http://adsabs.harvard.edu/abs/2006ApJS..167....1B} {167, 1}

\bibitem[\protect\citeauthoryear{{Bertelli}, {Bressan}, {Chiosi}, {Fagotto}  \&
  {Nasi}}{{Bertelli} et~al.}{1994}]{Bertelli94}
{Bertelli} G.,  {Bressan} A.,  {Chiosi} C.,  {Fagotto} F.,   {Nasi} E.,  1994,
  \aaps, \href {http://adsabs.harvard.edu/abs/1994A%26AS..106..275B} {106}

\bibitem[\protect\citeauthoryear{{Blanton}, {Eisenstein}, {Hogg}, {Schlegel},
  {Brinkmann}, {Quintero}, {Berlind}  \& {Wherry}}{{Blanton}
  et~al.}{2003}]{Blanton03}
{Blanton} M.~R.,  {Eisenstein} D.~J.,  {Hogg} D.~W.,  {Schlegel} D.~J.~S.,
  {Brinkmann} J.,  {Quintero} A.~D.,  {Berlind} A.,   {Wherry} N.,  2003, in
  American Astronomical Society Meeting Abstracts. p.~589

\bibitem[\protect\citeauthoryear{{Boselli}, {Gavazzi}  \& {Boissier}}{{Boselli}
  et~al.}{2006}]{Boselli06}
{Boselli} A.,  {Gavazzi} G.,   {Boissier} S.,  2006, in {Barret} D.,  {Casoli}
  F.,  {Lagache} G.,  {Lecavelier} A.,   {Pagani} L.,  eds, SF2A-2006: Semaine
  de l'Astrophysique Francaise. p.~317

\bibitem[\protect\citeauthoryear{{Brinchmann}, {Charlot}, {White}, {Tremonti},
  {Kauffmann}, {Heckman}  \& {Brinkmann}}{{Brinchmann}
  et~al.}{2004}]{Brinchmann04}
{Brinchmann} J.,  {Charlot} S.,  {White} S.~D.~M.,  {Tremonti} C.,  {Kauffmann}
  G.,  {Heckman} T.,   {Brinkmann} J.,  2004, \mn@doi [\mnras]
  {10.1111/j.1365-2966.2004.07881.x}, \href
  {http://adsabs.harvard.edu/abs/2004MNRAS.351.1151B} {351, 1151}

\bibitem[\protect\citeauthoryear{{Butcher} \& {Oemler}}{{Butcher} \&
  {Oemler}}{1984}]{Butcher84}
{Butcher} H.,  {Oemler} Jr. A.,  1984, \mn@doi [\apj] {10.1086/162519}, \href
  {http://adsabs.harvard.edu/abs/1984ApJ...285..426B} {285, 426}

\bibitem[\protect\citeauthoryear{{Calvi}, {Poggianti}  \& {Vulcani}}{{Calvi}
  et~al.}{2011}]{Calvi11}
{Calvi} R.,  {Poggianti} B.~M.,   {Vulcani} B.,  2011, \mn@doi [MNRAS]
  {10.1111/j.1365-2966.2011.19088.x}, \href
  {http://adsabs.harvard.edu/abs/2011MNRAS.416..727C} {416, 727}

\bibitem[\protect\citeauthoryear{{Calvi}, {Poggianti}, {Fasano}  \&
  {Vulcani}}{{Calvi} et~al.}{2012}]{Calvi12}
{Calvi} R.,  {Poggianti} B.~M.,  {Fasano} G.,   {Vulcani} B.,  2012, \mn@doi
  [\mnras] {10.1111/j.1745-3933.2011.01168.x}, \href
  {http://adsabs.harvard.edu/abs/2012MNRAS.419L..14C} {419, L14}

\bibitem[\protect\citeauthoryear{{Calvi}, {Poggianti}, {Vulcani}  \&
  {Fasano}}{{Calvi} et~al.}{2013}]{Calvi13}
{Calvi} R.,  {Poggianti} B.~M.,  {Vulcani} B.,   {Fasano} G.,  2013, \mn@doi
  [\mnras] {10.1093/mnras/stt667}, \href
  {http://adsabs.harvard.edu/abs/2013MNRAS.432.3141C} {432, 3141}

\bibitem[\protect\citeauthoryear{{Cava} et~al.,}{{Cava} et~al.}{2009}]{Cava09}
{Cava} A.,  et~al., 2009, \mn@doi [\aap] {10.1051/0004-6361:200810997}, \href
  {http://adsabs.harvard.edu/abs/2009A%26A...495..707C} {495, 707}

\bibitem[\protect\citeauthoryear{{Conselice}, {O'Neil}, {Gallagher}  \&
  {Wyse}}{{Conselice} et~al.}{2003}]{Conselice03}
{Conselice} C.~J.,  {O'Neil} K.,  {Gallagher} J.~S.,   {Wyse} R.~F.~G.,  2003,
  \mn@doi [\apj] {10.1086/375216}, \href
  {http://adsabs.harvard.edu/abs/2003ApJ...591..167C} {591, 167}

\bibitem[\protect\citeauthoryear{{Cowie}, {Songaila}, {Hu}  \& {Cohen}}{{Cowie}
  et~al.}{1996}]{Cowie96}
{Cowie} L.~L.,  {Songaila} A.,  {Hu} E.~M.,   {Cohen} J.~G.,  1996, \mn@doi
  [\aj] {10.1086/118058}, \href
  {http://adsabs.harvard.edu/abs/1996AJ....112..839C} {112, 839}

\bibitem[\protect\citeauthoryear{{Daddi} et~al.,}{{Daddi}
  et~al.}{2007}]{Daddi07}
{Daddi} E.,  et~al., 2007, \mn@doi [\apj] {10.1086/521818}, \href
  {http://adsabs.harvard.edu/abs/2007ApJ...670..156D} {670, 156}

\bibitem[\protect\citeauthoryear{{Deng}}{{Deng}}{2010}]{Deng10}
{Deng} X.-F.,  2010, \mn@doi [\apj] {10.1088/0004-637X/721/1/809}, \href
  {http://adsabs.harvard.edu/abs/2010ApJ...721..809D} {721, 809}

\bibitem[\protect\citeauthoryear{{Desai} et~al.,}{{Desai}
  et~al.}{2007}]{Desai07}
{Desai} V.,  et~al., 2007, \mn@doi [\apj] {10.1086/513310}, \href
  {http://adsabs.harvard.edu/abs/2007ApJ...660.1151D} {660, 1151}

\bibitem[\protect\citeauthoryear{{Dressler}, {Thompson}  \&
  {Shectman}}{{Dressler} et~al.}{1985}]{Dressler85}
{Dressler} A.,  {Thompson} I.~B.,   {Shectman} S.~A.,  1985, \mn@doi [\apj]
  {10.1086/162813}, \href {http://adsabs.harvard.edu/abs/1985ApJ...288..481D}
  {288, 481}

\bibitem[\protect\citeauthoryear{{Dressler} et~al.,}{{Dressler}
  et~al.}{1997}]{Dressler97}
{Dressler} A.,  et~al., 1997, \apj, \href
  {http://adsabs.harvard.edu/abs/1997ApJ...490..577D} {490, 577}

\bibitem[\protect\citeauthoryear{{Driver}}{{Driver}}{2004}]{Driver04}
{Driver} S.~P.,  2004, in American Astronomical Society Meeting Abstracts
  \#204. p.~781

\bibitem[\protect\citeauthoryear{{Driver}, {Liske}, {Cross}, {De Propris}  \&
  {Allen}}{{Driver} et~al.}{2005}]{Driver05}
{Driver} S.~P.,  {Liske} J.,  {Cross} N.~J.~G.,  {De Propris} R.,   {Allen}
  P.~D.,  2005, \mn@doi [\mnras] {10.1111/j.1365-2966.2005.08990.x}, \href
  {http://adsabs.harvard.edu/abs/2005MNRAS.360...81D} {360, 81}

\bibitem[\protect\citeauthoryear{{Driver} et~al.,}{{Driver}
  et~al.}{2011}]{Driver11}
{Driver} S.~P.,  et~al., 2011, \mn@doi [\mnras]
  {10.1111/j.1365-2966.2010.18188.x}, \href
  {http://adsabs.harvard.edu/abs/2011MNRAS.413..971D} {413, 971}

\bibitem[\protect\citeauthoryear{{Dutton}, {van den Bosch}  \&
  {Dekel}}{{Dutton} et~al.}{2010}]{Dutton10}
{Dutton} A.~A.,  {van den Bosch} F.~C.,   {Dekel} A.,  2010, \mn@doi [\mnras]
  {10.1111/j.1365-2966.2010.16620.x}, \href
  {http://adsabs.harvard.edu/abs/2010MNRAS.405.1690D} {405, 1690}

\bibitem[\protect\citeauthoryear{{Eke} et~al.,}{{Eke} et~al.}{2004}]{Eke04}
{Eke} V.~R.,  et~al., 2004, \mn@doi [\mnras]
  {10.1111/j.1365-2966.2004.07408.x}, \href
  {http://adsabs.harvard.edu/abs/2004MNRAS.348..866E} {348, 866}

\bibitem[\protect\citeauthoryear{{Elbaz} et~al.,}{{Elbaz}
  et~al.}{2007}]{Elbaz07}
{Elbaz} D.,  et~al., 2007, \mn@doi [\aap] {10.1051/0004-6361:20077525}, \href
  {http://adsabs.harvard.edu/abs/2007A%26A...468...33E} {468, 33}

\bibitem[\protect\citeauthoryear{{Fasano}, {Poggianti}, {Couch}, {Bettoni},
  {Kj{\ae}rgaard}  \& {Moles}}{{Fasano} et~al.}{2000}]{Fasano00}
{Fasano} G.,  {Poggianti} B.~M.,  {Couch} W.~J.,  {Bettoni} D.,
  {Kj{\ae}rgaard} P.,   {Moles} M.,  2000, \mn@doi [\apj] {10.1086/317047},
  \href {http://adsabs.harvard.edu/abs/2000ApJ...542..673F} {542, 673}

\bibitem[\protect\citeauthoryear{{Fasano} et~al.,}{{Fasano}
  et~al.}{2012}]{Fasano12}
{Fasano} G.,  et~al., 2012, \mn@doi [\mnras]
  {10.1111/j.1365-2966.2011.19798.x}, \href
  {http://adsabs.harvard.edu/abs/2012MNRAS.420..926F} {420, 926}

\bibitem[\protect\citeauthoryear{{Finlator}, {Dav{\'e}}, {Papovich}  \&
  {Hernquist}}{{Finlator} et~al.}{2006}]{Finlator06}
{Finlator} K.,  {Dav{\'e}} R.,  {Papovich} C.,   {Hernquist} L.,  2006, \mn@doi
  [\apj] {10.1086/499349}, \href
  {http://adsabs.harvard.edu/abs/2006ApJ...639..672F} {639, 672}

\bibitem[\protect\citeauthoryear{{Finlator}, {Dav{\'e}}  \&
  {Oppenheimer}}{{Finlator} et~al.}{2007}]{Finlator07}
{Finlator} K.,  {Dav{\'e}} R.,   {Oppenheimer} B.~D.,  2007, \mn@doi [\mnras]
  {10.1111/j.1365-2966.2007.11578.x}, \href
  {http://adsabs.harvard.edu/abs/2007MNRAS.376.1861F} {376, 1861}

\bibitem[\protect\citeauthoryear{{Finn} et~al.,}{{Finn} et~al.}{2005}]{Finn05}
{Finn} R.~A.,  et~al., 2005, \mn@doi [\apj] {10.1086/431642}, \href
  {http://adsabs.harvard.edu/abs/2005ApJ...630..206F} {630, 206}

\bibitem[\protect\citeauthoryear{{Flores} et~al.,}{{Flores}
  et~al.}{1999}]{Flores99}
{Flores} H.,  et~al., 1999, \mn@doi [\apj] {10.1086/307172}, \href
  {http://adsabs.harvard.edu/abs/1999ApJ...517..148F} {517, 148}

\bibitem[\protect\citeauthoryear{{Fossati} et~al.,}{{Fossati}
  et~al.}{2015}]{Fossati15}
{Fossati} M.,  et~al., 2015, \mn@doi [\mnras] {10.1093/mnras/stu2255}, \href
  {http://adsabs.harvard.edu/abs/2015MNRAS.446.2582F} {446, 2582}

\bibitem[\protect\citeauthoryear{{Fritz} et~al.,}{{Fritz}
  et~al.}{2007}]{Fritz07}
{Fritz} J.,  et~al., 2007, \mn@doi [\aap] {10.1051/0004-6361:20077097}, \href
  {http://adsabs.harvard.edu/abs/2007A%26A...470..137F} {470, 137}

\bibitem[\protect\citeauthoryear{{Fritz} et~al.,}{{Fritz}
  et~al.}{2011}]{Fritz11}
{Fritz} J.,  et~al., 2011, \mn@doi [\aap] {10.1051/0004-6361/201015214}, \href
  {http://adsabs.harvard.edu/abs/2011A%26A...526A..45F} {526, A45}

\bibitem[\protect\citeauthoryear{{Fritz} et~al.,}{{Fritz}
  et~al.}{2014}]{Fritz14}
{Fritz} J.,  et~al., 2014, \mn@doi [\aap] {10.1051/0004-6361/201323138}, \href
  {http://adsabs.harvard.edu/abs/2014A%26A...566A..32F} {566, A32}

\bibitem[\protect\citeauthoryear{{Gavazzi} \& {Jaffe}}{{Gavazzi} \&
  {Jaffe}}{1985}]{Gavazzi85}
{Gavazzi} G.,  {Jaffe} W.,  1985, \mn@doi [\apjl] {10.1086/184515}, \href
  {http://adsabs.harvard.edu/abs/1985ApJ...294L..89G} {294, L89}

\bibitem[\protect\citeauthoryear{{Gavazzi}, {Boselli}, {Pedotti}, {Gallazzi}
  \& {Carrasco}}{{Gavazzi} et~al.}{2002}]{Gavazzi02}
{Gavazzi} G.,  {Boselli} A.,  {Pedotti} P.,  {Gallazzi} A.,   {Carrasco} L.,
  2002, \mn@doi [\aap] {10.1051/0004-6361:20021403}, \href
  {http://adsabs.harvard.edu/abs/2002A%26A...396..449G} {396, 449}

\bibitem[\protect\citeauthoryear{{Gisler}}{{Gisler}}{1978}]{Gisler78}
{Gisler} G.~R.,  1978, \mn@doi [\mnras] {10.1093/mnras/183.4.633}, \href
  {http://adsabs.harvard.edu/abs/1978MNRAS.183..633G} {183, 633}

\bibitem[\protect\citeauthoryear{{G{\'o}mez} et~al.,}{{G{\'o}mez}
  et~al.}{2003}]{Gomez03}
{G{\'o}mez} P.~L.,  et~al., 2003, \mn@doi [\apj] {10.1086/345593}, \href
  {http://adsabs.harvard.edu/abs/2003ApJ...584..210G} {584, 210}

\bibitem[\protect\citeauthoryear{{Guglielmo}, {Poggianti}, {Moretti}, {Fritz},
  {Calvi}, {Vulcani}, {Fasano}  \& {Paccagnella}}{{Guglielmo}
  et~al.}{2015}]{Guglielmo15}
{Guglielmo} V.,  {Poggianti} B.~M.,  {Moretti} A.,  {Fritz} J.,  {Calvi} R.,
  {Vulcani} B.,  {Fasano} G.,   {Paccagnella} A.,  2015, \mn@doi [\mnras]
  {10.1093/mnras/stv757}, \href
  {http://adsabs.harvard.edu/abs/2015MNRAS.450.2749G} {450, 2749}

\bibitem[\protect\citeauthoryear{{Gunn} \& {Gott}}{{Gunn} \&
  {Gott}}{1972}]{Gunn72}
{Gunn} J.~E.,  {Gott} III J.~R.,  1972, \mn@doi [\apj] {10.1086/151605}, \href
  {http://adsabs.harvard.edu/abs/1972ApJ...176....1G} {176, 1}

\bibitem[\protect\citeauthoryear{{Haarsma}, {Partridge}, {Windhorst}  \&
  {Richards}}{{Haarsma} et~al.}{2000}]{Haarsma00}
{Haarsma} D.~B.,  {Partridge} R.~B.,  {Windhorst} R.~A.,   {Richards} E.~A.,
  2000, \mn@doi [\apj] {10.1086/317225}, \href
  {http://adsabs.harvard.edu/abs/2000ApJ...544..641H} {544, 641}

\bibitem[\protect\citeauthoryear{{Haines} et~al.,}{{Haines}
  et~al.}{2013}]{Haines13}
{Haines} C.~P.,  et~al., 2013, \mn@doi [\apj] {10.1088/0004-637X/775/2/126},
  \href {http://adsabs.harvard.edu/abs/2013ApJ...775..126H} {775, 126}

\bibitem[\protect\citeauthoryear{{Hashimoto}, {Oemler}, {Lin}  \&
  {Tucker}}{{Hashimoto} et~al.}{1998}]{Hashimoto98}
{Hashimoto} Y.,  {Oemler} Jr. A.,  {Lin} H.,   {Tucker} D.~L.,  1998, \apj,
  \href {http://adsabs.harvard.edu/abs/1998ApJ...499..589H} {499, 589}

\bibitem[\protect\citeauthoryear{{Head}, {Lucey}, {Hudson}  \& {Smith}}{{Head}
  et~al.}{2014}]{Head14}
{Head} J.~T.~C.~G.,  {Lucey} J.~R.,  {Hudson} M.~J.,   {Smith} R.~J.,  2014,
  \mn@doi [\mnras] {10.1093/mnras/stu325}, \href
  {http://adsabs.harvard.edu/abs/2014MNRAS.440.1690H} {440, 1690}

\bibitem[\protect\citeauthoryear{{Hogg}, {Cohen}, {Blandford}  \&
  {Pahre}}{{Hogg} et~al.}{1998}]{Hogg98}
{Hogg} D.~W.,  {Cohen} J.~G.,  {Blandford} R.,   {Pahre} M.~A.,  1998, \mn@doi
  [\apj] {10.1086/306122}, \href
  {http://adsabs.harvard.edu/abs/1998ApJ...504..622H} {504, 622}

\bibitem[\protect\citeauthoryear{{Hopkins}}{{Hopkins}}{2004}]{Hopkins04}
{Hopkins} A.~M.,  2004, \mn@doi [\apj] {10.1086/424032}, \href
  {http://adsabs.harvard.edu/abs/2004ApJ...615..209H} {615, 209}

\bibitem[\protect\citeauthoryear{{Hudson}, {Stevenson}, {Smith}, {Wegner},
  {Lucey}  \& {Simard}}{{Hudson} et~al.}{2010}]{Hudson10}
{Hudson} M.~J.,  {Stevenson} J.~B.,  {Smith} R.~J.,  {Wegner} G.~A.,  {Lucey}
  J.~R.,   {Simard} L.,  2010, \mn@doi [\mnras]
  {10.1111/j.1365-2966.2010.17318.x}, \href
  {http://adsabs.harvard.edu/abs/2010MNRAS.409..405H} {409, 405}

\bibitem[\protect\citeauthoryear{{Jacoby}, {Hunter}  \& {Christian}}{{Jacoby}
  et~al.}{1984}]{Jacoby84}
{Jacoby} G.~H.,  {Hunter} D.~A.,   {Christian} C.~A.,  1984, \mn@doi [\apjs]
  {10.1086/190983}, \href {http://adsabs.harvard.edu/abs/1984ApJS...56..257J}
  {56, 257}

\bibitem[\protect\citeauthoryear{{Kannappan} \& {Gawiser}}{{Kannappan} \&
  {Gawiser}}{2007}]{Kannappan07}
{Kannappan} S.~J.,  {Gawiser} E.,  2007, \mn@doi [\apjl] {10.1086/512974},
  \href {http://adsabs.harvard.edu/abs/2007ApJ...657L...5K} {657, L5}

\bibitem[\protect\citeauthoryear{{Karim} et~al.,}{{Karim}
  et~al.}{2011}]{Karim11}
{Karim} A.,  et~al., 2011, \mn@doi [\apj] {10.1088/0004-637X/730/2/61}, \href
  {http://adsabs.harvard.edu/abs/2011ApJ...730...61K} {730, 61}

\bibitem[\protect\citeauthoryear{{Kauffmann} et~al.,}{{Kauffmann}
  et~al.}{2003}]{Kauffmann03}
{Kauffmann} G.,  et~al., 2003, \mn@doi [\mnras]
  {10.1046/j.1365-8711.2003.06292.x}, \href
  {http://adsabs.harvard.edu/abs/2003MNRAS.341...54K} {341, 54}

\bibitem[\protect\citeauthoryear{{Kauffmann}, {White}, {Heckman}, {M{\'e}nard},
  {Brinchmann}, {Charlot}, {Tremonti}  \& {Brinkmann}}{{Kauffmann}
  et~al.}{2004}]{Kauffmann04}
{Kauffmann} G.,  {White} S.~D.~M.,  {Heckman} T.~M.,  {M{\'e}nard} B.,
  {Brinchmann} J.,  {Charlot} S.,  {Tremonti} C.,   {Brinkmann} J.,  2004,
  \mn@doi [\mnras] {10.1111/j.1365-2966.2004.08117.x}, \href
  {http://adsabs.harvard.edu/abs/2004MNRAS.353..713K} {353, 713}

\bibitem[\protect\citeauthoryear{{Kenney}, {Geha}, {J{\'a}chym}, {Crowl},
  {Dague}, {Chung}, {van Gorkom}  \& {Vollmer}}{{Kenney}
  et~al.}{2014}]{Kenney14}
{Kenney} J.~D.~P.,  {Geha} M.,  {J{\'a}chym} P.,  {Crowl} H.~H.,  {Dague} W.,
  {Chung} A.,  {van Gorkom} J.,   {Vollmer} B.,  2014, \mn@doi [\apj]
  {10.1088/0004-637X/780/2/119}, \href
  {http://adsabs.harvard.edu/abs/2014ApJ...780..119K} {780, 119}

\bibitem[\protect\citeauthoryear{{Kennicutt}}{{Kennicutt}}{1983}]{Kennicutt83}
{Kennicutt} Jr. R.~C.,  1983, \mn@doi [\apj] {10.1086/161261}, \href
  {http://adsabs.harvard.edu/abs/1983ApJ...272...54K} {272, 54}

\bibitem[\protect\citeauthoryear{{Kennicutt}}{{Kennicutt}}{1992}]{Kennicutt92}
{Kennicutt} Jr. R.~C.,  1992, \mn@doi [\apj] {10.1086/171154}, \href
  {http://adsabs.harvard.edu/abs/1992ApJ...388..310K} {388, 310}

\bibitem[\protect\citeauthoryear{{Kennicutt}, {Bothun}  \&
  {Schommer}}{{Kennicutt} et~al.}{1984}]{Kennicutt84}
{Kennicutt} Jr. R.~C.,  {Bothun} G.~D.,   {Schommer} R.~A.,  1984, \mn@doi
  [\aj] {10.1086/113625}, \href
  {http://adsabs.harvard.edu/abs/1984AJ.....89.1279K} {89, 1279}

\bibitem[\protect\citeauthoryear{{Kewley}, {Geller}  \& {Barton}}{{Kewley}
  et~al.}{2006}]{Kewley06}
{Kewley} L.~J.,  {Geller} M.~J.,   {Barton} E.~J.,  2006, \mn@doi [\aj]
  {10.1086/500295}, \href {http://adsabs.harvard.edu/abs/2006AJ....131.2004K}
  {131, 2004}

\bibitem[\protect\citeauthoryear{{Kravtsov}, {Gnedin}  \& {Klypin}}{{Kravtsov}
  et~al.}{2004}]{Kravtsov04}
{Kravtsov} A.~V.,  {Gnedin} O.~Y.,   {Klypin} A.~A.,  2004, \mn@doi [\apj]
  {10.1086/421322}, \href {http://adsabs.harvard.edu/abs/2004ApJ...609..482K}
  {609, 482}

\bibitem[\protect\citeauthoryear{{Kroupa}}{{Kroupa}}{2001}]{Kroupa01}
{Kroupa} P.,  2001, \mn@doi [\mnras] {10.1046/j.1365-8711.2001.04022.x}, \href
  {http://adsabs.harvard.edu/abs/2001MNRAS.322..231K} {322, 231}

\bibitem[\protect\citeauthoryear{{Lara-L{\'o}pez} et~al.,}{{Lara-L{\'o}pez}
  et~al.}{2013}]{LaraLopez13}
{Lara-L{\'o}pez} M.~A.,  et~al., 2013, \mn@doi [\mnras]
  {10.1093/mnras/stt1031}, \href
  {http://adsabs.harvard.edu/abs/2013MNRAS.434..451L} {434, 451}

\bibitem[\protect\citeauthoryear{{Lewis} et~al.,}{{Lewis}
  et~al.}{2002}]{Lewis02}
{Lewis} I.,  et~al., 2002, \mn@doi [\mnras] {10.1046/j.1365-8711.2002.05558.x},
  \href {http://adsabs.harvard.edu/abs/2002MNRAS.334..673L} {334, 673}

\bibitem[\protect\citeauthoryear{{Lilly}, {Le Fevre}, {Hammer}  \&
  {Crampton}}{{Lilly} et~al.}{1996}]{Lilly96}
{Lilly} S.~J.,  {Le Fevre} O.,  {Hammer} F.,   {Crampton} D.,  1996, \mn@doi
  [\apjl] {10.1086/309975}, \href
  {http://adsabs.harvard.edu/abs/1996ApJ...460L...1L} {460, L1}

\bibitem[\protect\citeauthoryear{{Lilly} et~al.,}{{Lilly}
  et~al.}{2007}]{Lilly07}
{Lilly} S.~J.,  et~al., 2007, \mn@doi [\apjs] {10.1086/516589}, \href
  {http://adsabs.harvard.edu/abs/2007ApJS..172...70L} {172, 70}

\bibitem[\protect\citeauthoryear{{Lin} et~al.,}{{Lin} et~al.}{2014}]{Lin14}
{Lin} L.,  et~al., 2014, \mn@doi [\apj] {10.1088/0004-637X/782/1/33}, \href
  {http://adsabs.harvard.edu/abs/2014ApJ...782...33L} {782, 33}

\bibitem[\protect\citeauthoryear{{Liske}, {Lemon}, {Driver}, {Cross}  \&
  {Couch}}{{Liske} et~al.}{2003}]{Liske03}
{Liske} J.,  {Lemon} D.~J.,  {Driver} S.~P.,  {Cross} N.~J.~G.,   {Couch}
  W.~J.,  2003, \mn@doi [\mnras] {10.1046/j.1365-8711.2003.06826.x}, \href
  {http://adsabs.harvard.edu/abs/2003MNRAS.344..307L} {344, 307}

\bibitem[\protect\citeauthoryear{{Madau}, {Ferguson}, {Dickinson},
  {Giavalisco}, {Steidel}  \& {Fruchter}}{{Madau} et~al.}{1996}]{Madau96}
{Madau} P.,  {Ferguson} H.~C.,  {Dickinson} M.~E.,  {Giavalisco} M.,  {Steidel}
  C.~C.,   {Fruchter} A.,  1996, \mn@doi [\mnras] {10.1093/mnras/283.4.1388},
  \href {http://adsabs.harvard.edu/abs/1996MNRAS.283.1388M} {283, 1388}

\bibitem[\protect\citeauthoryear{{Madau}, {Pozzetti}  \& {Dickinson}}{{Madau}
  et~al.}{1998}]{Madau98}
{Madau} P.,  {Pozzetti} L.,   {Dickinson} M.,  1998, \mn@doi [\apj]
  {10.1086/305523}, \href {http://adsabs.harvard.edu/abs/1998ApJ...498..106M}
  {498, 106}

\bibitem[\protect\citeauthoryear{{Mart{\'{\i}}nez}, {Coenda}  \&
  {Muriel}}{{Mart{\'{\i}}nez} et~al.}{2008}]{Martinez08}
{Mart{\'{\i}}nez} H.~J.,  {Coenda} V.,   {Muriel} H.,  2008, \mn@doi [\mnras]
  {10.1111/j.1365-2966.2008.13929.x}, \href
  {http://adsabs.harvard.edu/abs/2008MNRAS.391..585M} {391, 585}

\bibitem[\protect\citeauthoryear{{McGee}, {Balogh}, {Wilman}, {Bower},
  {Mulchaey}, {Parker}  \& {Oemler}}{{McGee} et~al.}{2011}]{McGee11}
{McGee} S.~L.,  {Balogh} M.~L.,  {Wilman} D.~J.,  {Bower} R.~G.,  {Mulchaey}
  J.~S.,  {Parker} L.~C.,   {Oemler} A.,  2011, \mn@doi [\mnras]
  {10.1111/j.1365-2966.2010.18189.x}, \href
  {http://adsabs.harvard.edu/abs/2011MNRAS.413..996M} {413, 996}

\bibitem[\protect\citeauthoryear{{Moore}, {Katz}, {Lake}, {Dressler}  \&
  {Oemler}}{{Moore} et~al.}{1996}]{Moore96}
{Moore} B.,  {Katz} N.,  {Lake} G.,  {Dressler} A.,   {Oemler} A.,  1996,
  \mn@doi [\nat] {10.1038/379613a0}, \href
  {http://adsabs.harvard.edu/abs/1996Natur.379..613M} {379, 613}

\bibitem[\protect\citeauthoryear{{Moss} \& {Whittle}}{{Moss} \&
  {Whittle}}{1993}]{Moss93}
{Moss} C.,  {Whittle} M.,  1993, \mn@doi [\apjl] {10.1086/186795}, \href
  {http://adsabs.harvard.edu/abs/1993ApJ...407L..17M} {407, L17}

\bibitem[\protect\citeauthoryear{{Muldrew} et~al.,}{{Muldrew}
  et~al.}{2012}]{Muldrew12}
{Muldrew} S.~I.,  et~al., 2012, \mn@doi [\mnras]
  {10.1111/j.1365-2966.2011.19922.x}, \href
  {http://adsabs.harvard.edu/abs/2012MNRAS.419.2670M} {419, 2670}

\bibitem[\protect\citeauthoryear{{Noeske} et~al.,}{{Noeske}
  et~al.}{2007}]{Noeske07a}
{Noeske} K.~G.,  et~al., 2007, \mn@doi [\apjl] {10.1086/517926}, \href
  {http://adsabs.harvard.edu/abs/2007ApJ...660L..43N} {660, L43}

\bibitem[\protect\citeauthoryear{{Oemler}, {Abramson}, {Gladders}, {Dressler},
  {Poggianti}  \& {Vulcani}}{{Oemler} et~al.}{2016}]{Oemler16}
{Oemler} Jr A.,  {Abramson} L.~E.,  {Gladders} M.~D.,  {Dressler} A.,
  {Poggianti} B.~M.,   {Vulcani} B.,  2016, preprint, \href
  {http://adsabs.harvard.edu/abs/2016arXiv161105932O} {} (\mn@eprint {arXiv}
  {1611.05932})

\bibitem[\protect\citeauthoryear{{Paccagnella} et~al.,}{{Paccagnella}
  et~al.}{2016}]{Paccagnella16}
{Paccagnella} A.,  et~al., 2016, \mn@doi [\apjl] {10.3847/2041-8205/816/2/L25},
  \href {http://adsabs.harvard.edu/abs/2016ApJ...816L..25P} {816, L25}

\bibitem[\protect\citeauthoryear{{Paccagnella}, {Vulcani}, {Poggianti},
  {Moretti}, {Fritz}  \& {Fasano}}{{Paccagnella}
  et~al.}{2018}]{Paccagnella2018}
{Paccagnella} A.,  {Vulcani} B.,  {Poggianti} B.~M.,  {Moretti} A.,  {Fritz}
  J.,   {Fasano} G.,  2018, preprint, \href
  {http://adsabs.harvard.edu/abs/2018arXiv180511475P} {} (\mn@eprint {arXiv}
  {1805.11475})

\bibitem[\protect\citeauthoryear{{Pannella} et~al.,}{{Pannella}
  et~al.}{2009}]{Pannella09}
{Pannella} M.,  et~al., 2009, \mn@doi [\apjl] {10.1088/0004-637X/698/2/L116},
  \href {http://adsabs.harvard.edu/abs/2009ApJ...698L.116P} {698, L116}

\bibitem[\protect\citeauthoryear{{Patel}, {Holden}, {Kelson}, {Illingworth}  \&
  {Franx}}{{Patel} et~al.}{2009}]{Patel09}
{Patel} S.~G.,  {Holden} B.~P.,  {Kelson} D.~D.,  {Illingworth} G.~D.,
  {Franx} M.,  2009, \mn@doi [\apjl] {10.1088/0004-637X/705/1/L67}, \href
  {http://adsabs.harvard.edu/abs/2009ApJ...705L..67P} {705, L67}

\bibitem[\protect\citeauthoryear{{Peng} et~al.,}{{Peng} et~al.}{2010}]{Peng10}
{Peng} Y.-j.,  et~al., 2010, \mn@doi [\apj] {10.1088/0004-637X/721/1/193},
  \href {http://adsabs.harvard.edu/abs/2010ApJ...721..193P} {721, 193}

\bibitem[\protect\citeauthoryear{{Poggianti} et~al.,}{{Poggianti}
  et~al.}{2009}]{Poggianti09}
{Poggianti} B.~M.,  et~al., 2009, \mn@doi [\apjl]
  {10.1088/0004-637X/697/2/L137}, \href
  {http://adsabs.harvard.edu/abs/2009ApJ...697L.137P} {697, L137}

\bibitem[\protect\citeauthoryear{{Poggianti}, {Moretti}, {Calvi}, {D'Onofrio},
  {Valentinuzzi}, {Fritz}  \& {Renzini}}{{Poggianti}
  et~al.}{2013}]{Poggianti13}
{Poggianti} B.~M.,  {Moretti} A.,  {Calvi} R.,  {D'Onofrio} M.,  {Valentinuzzi}
  T.,  {Fritz} J.,   {Renzini} A.,  2013, \mn@doi [\apj]
  {10.1088/0004-637X/777/2/125}, \href
  {http://adsabs.harvard.edu/abs/2013ApJ...777..125P} {777, 125}

\bibitem[\protect\citeauthoryear{{Postman} et~al.,}{{Postman}
  et~al.}{2005}]{Postman05}
{Postman} M.,  et~al., 2005, \mn@doi [\apj] {10.1086/428881}, \href
  {http://adsabs.harvard.edu/abs/2005ApJ...623..721P} {623, 721}

\bibitem[\protect\citeauthoryear{{Roberts} \& {Haynes}}{{Roberts} \&
  {Haynes}}{1994}]{RobertsHaynes94}
{Roberts} M.~S.,  {Haynes} M.,  1994, in {Meylan} G.,  {Prugniel} P.,  eds,
  European Southern Observatory Conference and Workshop Proceedings Vol. 49,
  European Southern Observatory Conference and Workshop Proceedings. p.~197

\bibitem[\protect\citeauthoryear{{Rodr{\'{\i}}guez-Puebla}, {Primack},
  {Behroozi}  \& {Faber}}{{Rodr{\'{\i}}guez-Puebla}
  et~al.}{2016}]{RodriguezPuebla16}
{Rodr{\'{\i}}guez-Puebla} A.,  {Primack} J.~R.,  {Behroozi} P.,   {Faber}
  S.~M.,  2016, \mn@doi [\mnras] {10.1093/mnras/stv2513}, \href
  {http://adsabs.harvard.edu/abs/2016MNRAS.455.2592R} {455, 2592}

\bibitem[\protect\citeauthoryear{{Salpeter}}{{Salpeter}}{1955}]{Salpeter55}
{Salpeter} E.~E.,  1955, \mn@doi [\apj] {10.1086/145971}, \href
  {http://adsabs.harvard.edu/abs/1955ApJ...121..161S} {121, 161}

\bibitem[\protect\citeauthoryear{{S{\'a}nchez-Bl{\'a}zquez}}{{S{\'a}nchez-Bl{\'a}zquez}}{2004}]{Sanchez04}
{S{\'a}nchez-Bl{\'a}zquez} P.,  2004, PhD thesis, Universidad Complutense de
  Madrid, Spain

\bibitem[\protect\citeauthoryear{{S{\'a}nchez-Bl{\'a}zquez}
  et~al.,}{{S{\'a}nchez-Bl{\'a}zquez} et~al.}{2006}]{Sanchez06}
{S{\'a}nchez-Bl{\'a}zquez} P.,  et~al., 2006, \mn@doi [\mnras]
  {10.1111/j.1365-2966.2006.10699.x}, \href
  {http://adsabs.harvard.edu/abs/2006MNRAS.371..703S} {371, 703}

\bibitem[\protect\citeauthoryear{{Smith}, {Kneib}, {Smail}, {Mazzotta},
  {Ebeling}  \& {Czoske}}{{Smith} et~al.}{2005}]{Smith05}
{Smith} G.~P.,  {Kneib} J.-P.,  {Smail} I.,  {Mazzotta} P.,  {Ebeling} H.,
  {Czoske} O.,  2005, \mn@doi [\mnras] {10.1111/j.1365-2966.2005.08911.x},
  \href {http://adsabs.harvard.edu/abs/2005MNRAS.359..417S} {359, 417}

\bibitem[\protect\citeauthoryear{{Sobral}, {Stroe}, {Dawson}, {Wittman}, {Jee},
  {R{\"o}ttgering}, {van Weeren}  \& {Br{\"u}ggen}}{{Sobral}
  et~al.}{2015}]{Sobral15}
{Sobral} D.,  {Stroe} A.,  {Dawson} W.~A.,  {Wittman} D.,  {Jee} M.~J.,
  {R{\"o}ttgering} H.,  {van Weeren} R.~J.,   {Br{\"u}ggen} M.,  2015, \mn@doi
  [\mnras] {10.1093/mnras/stv521}, \href
  {http://adsabs.harvard.edu/abs/2015MNRAS.450..630S} {450, 630}

\bibitem[\protect\citeauthoryear{{Sparre} et~al.,}{{Sparre}
  et~al.}{2015}]{Sparre15}
{Sparre} M.,  et~al., 2015, \mn@doi [\mnras] {10.1093/mnras/stu2713}, \href
  {http://adsabs.harvard.edu/abs/2015MNRAS.447.3548S} {447, 3548}

\bibitem[\protect\citeauthoryear{{Springel}}{{Springel}}{2005}]{Springel05}
{Springel} V.,  2005, \mn@doi [\mnras] {10.1111/j.1365-2966.2005.09655.x},
  \href {http://adsabs.harvard.edu/abs/2005MNRAS.364.1105S} {364, 1105}

\bibitem[\protect\citeauthoryear{{Strateva} et~al.,}{{Strateva}
  et~al.}{2001}]{Strateva01}
{Strateva} I.,  et~al., 2001, \mn@doi [\aj] {10.1086/323301}, \href
  {http://adsabs.harvard.edu/abs/2001AJ....122.1861S} {122, 1861}

\bibitem[\protect\citeauthoryear{{Stroe} et~al.,}{{Stroe}
  et~al.}{2015}]{Stroe15}
{Stroe} A.,  et~al., 2015, \mn@doi [\mnras] {10.1093/mnras/stu2519}, \href
  {http://adsabs.harvard.edu/abs/2015MNRAS.450..646S} {450, 646}

\bibitem[\protect\citeauthoryear{{Treu}, {Ellis}, {Kneib}, {Dressler}, {Smail},
  {Czoske}, {Oemler}  \& {Natarajan}}{{Treu} et~al.}{2003}]{Treu03}
{Treu} T.,  {Ellis} R.~S.,  {Kneib} J.-P.,  {Dressler} A.,  {Smail} I.,
  {Czoske} O.,  {Oemler} A.,   {Natarajan} P.,  2003, \mn@doi [\apj]
  {10.1086/375314}, \href {http://adsabs.harvard.edu/abs/2003ApJ...591...53T}
  {591, 53}

\bibitem[\protect\citeauthoryear{{Valentinuzzi} et~al.,}{{Valentinuzzi}
  et~al.}{2011}]{Valentinuzzi11}
{Valentinuzzi} T.,  et~al., 2011, \mn@doi [\aap] {10.1051/0004-6361/201117522},
  \href {http://adsabs.harvard.edu/abs/2011A%26A...536A..34V} {536, A34}

\bibitem[\protect\citeauthoryear{{Vulcani}, {Poggianti}, {Finn}, {Rudnick},
  {Desai}  \& {Bamford}}{{Vulcani} et~al.}{2010}]{Vulcani10}
{Vulcani} B.,  {Poggianti} B.~M.,  {Finn} R.~A.,  {Rudnick} G.,  {Desai} V.,
  {Bamford} S.,  2010, \mn@doi [\apjl] {10.1088/2041-8205/710/1/L1}, \href
  {http://adsabs.harvard.edu/abs/2010ApJ...710L...1V} {710, L1}

\bibitem[\protect\citeauthoryear{{Vulcani} et~al.,}{{Vulcani}
  et~al.}{2012}]{Vulcani12}
{Vulcani} B.,  et~al., 2012, \mn@doi [\mnras]
  {10.1111/j.1365-2966.2011.20135.x}, \href
  {http://adsabs.harvard.edu/abs/2012MNRAS.420.1481V} {420, 1481}

\bibitem[\protect\citeauthoryear{{Vulcani}, {Poggianti}, {Fritz}, {Fasano},
  {Moretti}, {Calvi}  \& {Paccagnella}}{{Vulcani} et~al.}{2015}]{Vulcani15}
{Vulcani} B.,  {Poggianti} B.~M.,  {Fritz} J.,  {Fasano} G.,  {Moretti} A.,
  {Calvi} R.,   {Paccagnella} A.,  2015, \mn@doi [\apj]
  {10.1088/0004-637X/798/1/52}, \href
  {http://adsabs.harvard.edu/abs/2015ApJ...798...52V} {798, 52}

\bibitem[\protect\citeauthoryear{{Willett} et~al.,}{{Willett}
  et~al.}{2015}]{Willett15}
{Willett} K.~W.,  et~al., 2015, \mn@doi [\mnras] {10.1093/mnras/stv307}, \href
  {http://adsabs.harvard.edu/abs/2015MNRAS.449..820W} {449, 820}

\bibitem[\protect\citeauthoryear{{Wuyts} et~al.,}{{Wuyts}
  et~al.}{2011}]{Wuyts11}
{Wuyts} S.,  et~al., 2011, \mn@doi [\apj] {10.1088/0004-637X/742/2/96}, \href
  {http://adsabs.harvard.edu/abs/2011ApJ...742...96W} {742, 96}

\bibitem[\protect\citeauthoryear{{Yang}, {Mo}, {van den Bosch}, {Pasquali},
  {Li}  \& {Barden}}{{Yang} et~al.}{2007}]{Yang07}
{Yang} X.,  {Mo} H.~J.,  {van den Bosch} F.~C.,  {Pasquali} A.,  {Li} C.,
  {Barden} M.,  2007, \mn@doi [\apj] {10.1086/522027}, \href
  {http://adsabs.harvard.edu/abs/2007ApJ...671..153Y} {671, 153}

\bibitem[\protect\citeauthoryear{{Yang}, {Mo}  \& {van den Bosch}}{{Yang}
  et~al.}{2008}]{Yang08}
{Yang} X.,  {Mo} H.~J.,   {van den Bosch} F.~C.,  2008, \mn@doi [\apj]
  {10.1086/528954}, \href {http://adsabs.harvard.edu/abs/2008ApJ...676..248Y}
  {676, 248}

\bibitem[\protect\citeauthoryear{{York} et~al.,}{{York} et~al.}{2000}]{York00}
{York} D.~G.,  et~al., 2000, \mn@doi [\aj] {10.1086/301513}, \href
  {http://adsabs.harvard.edu/abs/2000AJ....120.1579Y} {120, 1579}

\bibitem[\protect\citeauthoryear{{Yuan}, {Zhao}, {Yang}, {Wen}  \&
  {Zhou}}{{Yuan} et~al.}{2005}]{Yuan05}
{Yuan} Q.,  {Zhao} L.,  {Yang} Y.,  {Wen} Z.,   {Zhou} X.,  2005, \mn@doi [\aj]
  {10.1086/497390}, \href {http://adsabs.harvard.edu/abs/2005AJ....130.2559Y}
  {130, 2559}

\bibitem[\protect\citeauthoryear{{Zamojski} et~al.,}{{Zamojski}
  et~al.}{2007}]{Zamojski07}
{Zamojski} M.~A.,  et~al., 2007, \mn@doi [\apjs] {10.1086/516593}, \href
  {http://adsabs.harvard.edu/abs/2007ApJS..172..468Z} {172, 468}

\bibitem[\protect\citeauthoryear{{Zheng}, {Bell}, {Papovich}, {Wolf},
  {Meisenheimer}, {Rix}, {Rieke}  \& {Somerville}}{{Zheng}
  et~al.}{2007}]{Zheng07}
{Zheng} X.~Z.,  {Bell} E.~F.,  {Papovich} C.,  {Wolf} C.,  {Meisenheimer} K.,
  {Rix} H.-W.,  {Rieke} G.~H.,   {Somerville} R.,  2007, \mn@doi [\apjl]
  {10.1086/518690}, \href {http://adsabs.harvard.edu/abs/2007ApJ...661L..41Z}
  {661, L41}

\bibitem[\protect\citeauthoryear{{Zolotov} et~al.,}{{Zolotov}
  et~al.}{2012}]{Zolotov12}
{Zolotov} A.,  et~al., 2012, \mn@doi [\apj] {10.1088/0004-637X/761/1/71}, \href
  {http://adsabs.harvard.edu/abs/2012ApJ...761...71Z} {761, 71}

\bibitem[\protect\citeauthoryear{{van den Bosch}}{{van den
  Bosch}}{2002}]{Vandenbosch02}
{van den Bosch} F.~C.,  2002, \mn@doi [\mnras]
  {10.1046/j.1365-8711.2002.05171.x}, \href
  {http://adsabs.harvard.edu/abs/2002MNRAS.331...98V} {331, 98}

\bibitem[\protect\citeauthoryear{{von der Linden}, {Wild}, {Kauffmann}, {White}
   \& {Weinmann}}{{von der Linden} et~al.}{2010}]{VonderLinden10}
{von der Linden} A.,  {Wild} V.,  {Kauffmann} G.,  {White} S.~D.~M.,
  {Weinmann} S.,  2010, \mn@doi [\mnras] {10.1111/j.1365-2966.2010.16375.x},
  \href {http://adsabs.harvard.edu/abs/2010MNRAS.404.1231V} {404, 1231}

\makeatother
\end{thebibliography}

\appendix  
\section{SFR-M$\ast$ and SSFR-$M_\ast$ relations as a function of morphology using the local density and halo mass parameterizations}\label{appendix}

In this Appendix we present the SFR- and SSFR- $M_\ast$ relations for  ellipticals, S0s and late-type galaxies in at different local densities and at different halo masses. 

\begin{figure*}
\centering
\includegraphics[scale=0.55]{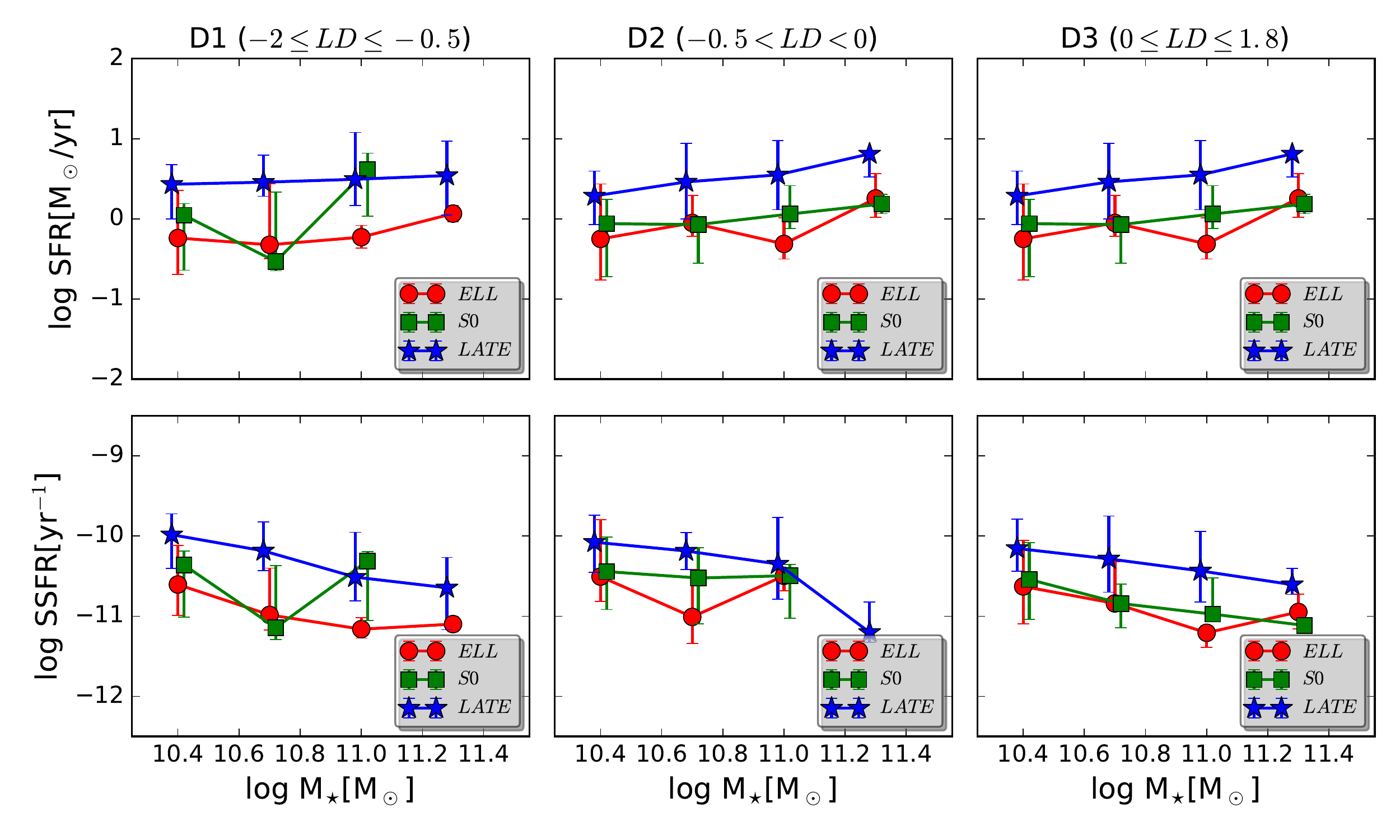}
\caption{Comparison between the SFR- and SSFR- $M_{\star}$ relations for ellipticals, S0s and late-type galaxies fixing the bin of local density. In each panel the red circle points are the medians for ellipticals, the green square points for S0s and the blue star points for late-types. Error bars represent the 25$^{th}$ and 75$^{th}$ percentiles. }\label{SFR_M_10}
\end{figure*}

\begin{figure*}
\centering
\includegraphics[scale=0.65]{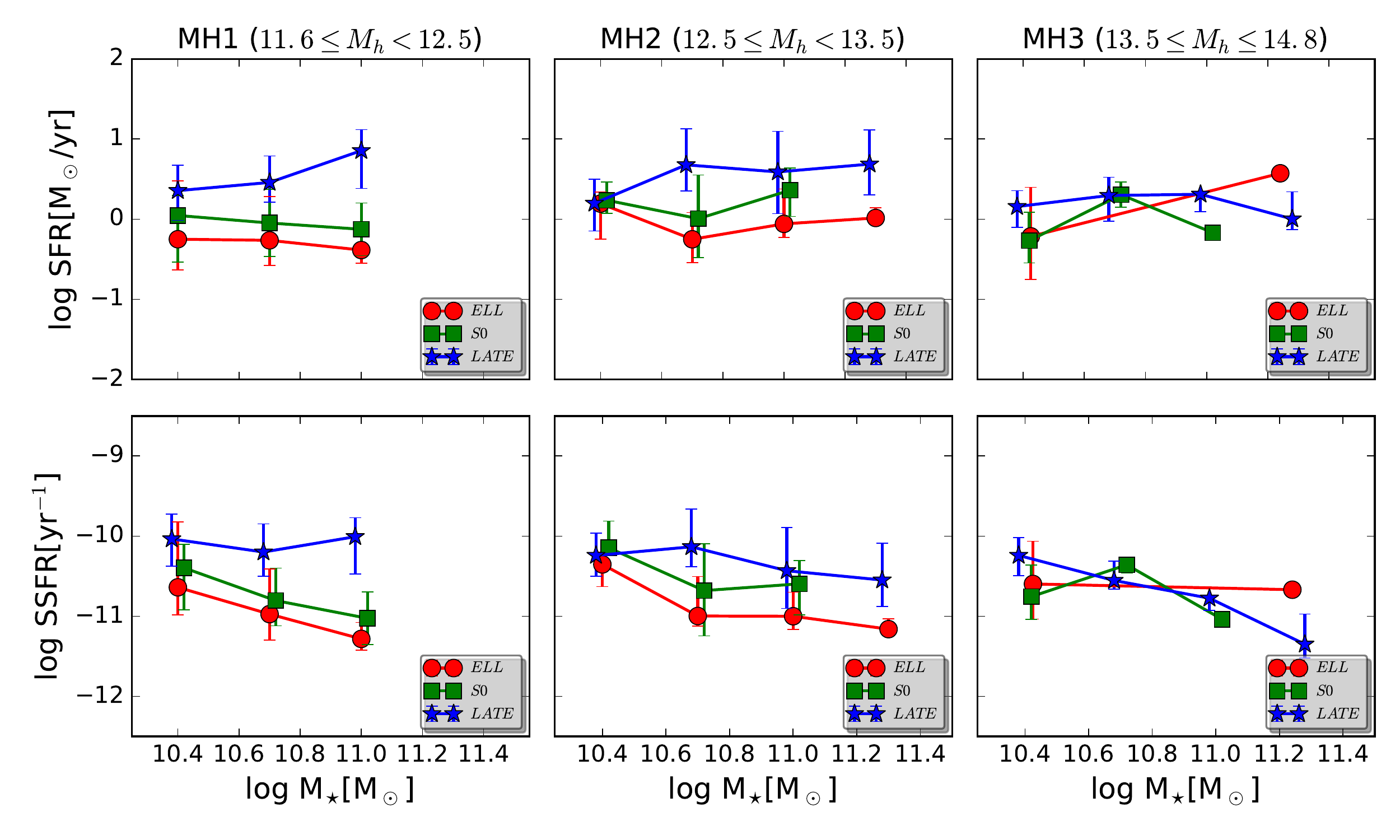}
\caption{Comparison between the SFR- and SSFR- $M_{\star}$ relations for ellipticals, S0s and late-type galaxies fixing the bin of halo mass. In each panel the red circle points are the medians for ellipticals, the green squares
points the medians for S0s and the blue star points the medians for late-types. Error bars represent the 25$^{th}$ and 75$^{th}$ percentiles.}\label{SFR_M_12}
\end{figure*}

In agreement to what presented in the main text for the global environment, fixing either  the local density (Figure \ref{SFR_M_10}) or the halo mass (Figure \ref{SFR_M_12}), we find a variation of the median SFR and SSFR with morphology: at any given mass values generally increase from ellipticals, to S0s, to late-types. Even though errors are quite large, results are solid. 

\begin{figure*}
\centering
\includegraphics[scale=0.65]{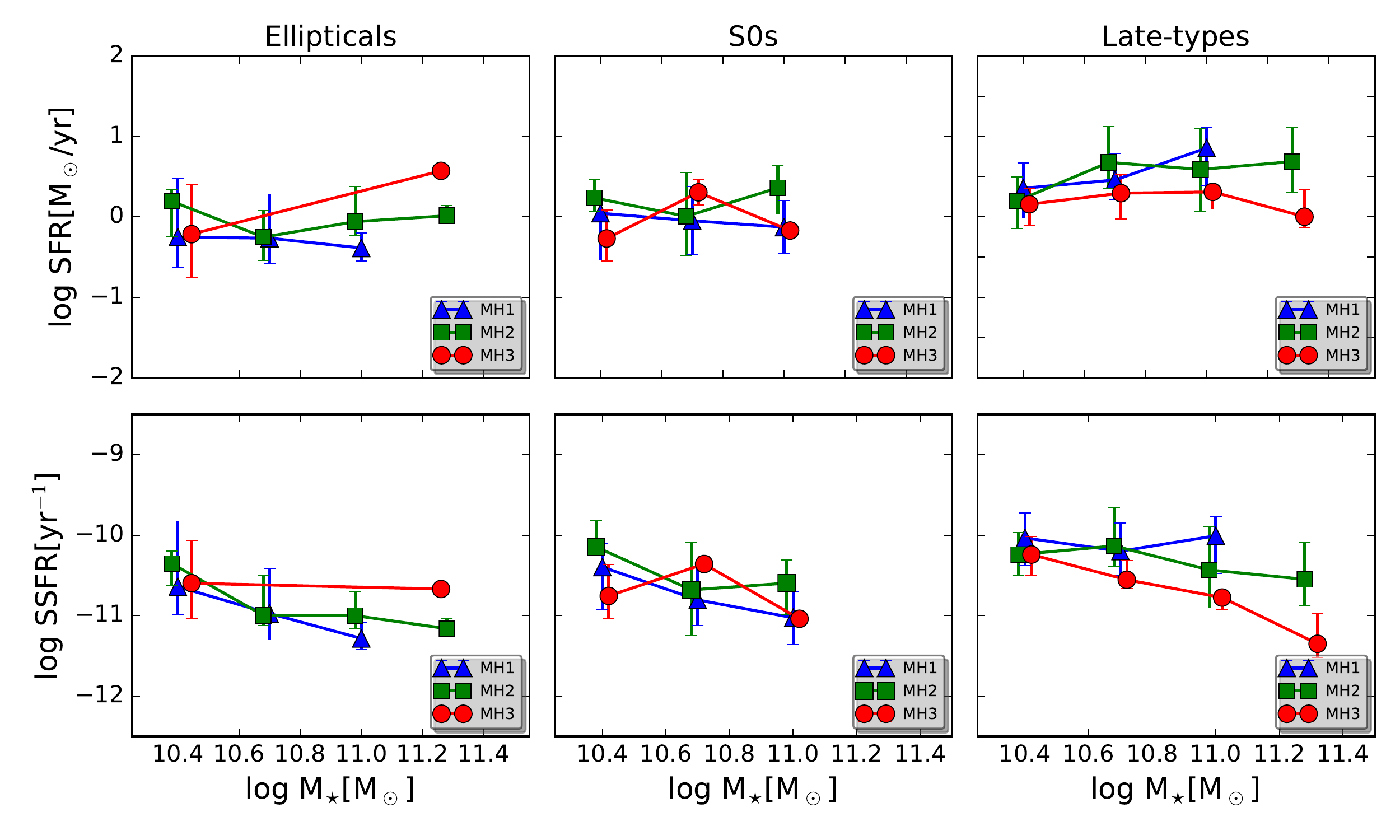}
\caption{Comparison between the SFR- and SSFR- $M_{\star}$ relations in different bins of halo mass fixing the morphological type: ellipticals (left panels), S0s (central panels) and late-type (right panels). In each panel the blue triangle points are the medians in the bin $11.7\leq M_{h}<12.5$, the green square
points the medians in the bin $12.5\leq M_{h}<13.5$ and the red circle points the medians in the bin $13.5\leq M_{h}\leq 14.8$. Error bars represent the 25$^{th}$ and 75$^{th}$ percentiles.}\label{SFR_M_11}
\end{figure*}

In contrast, fixing the morphology, 
Figure \ref{SFR_M_11} evidences almost no variation across the environments.

\label{lastpage}
\end{document}